\documentclass[final,twocolumn]{IEEEtran}

\usepackage{amsmath,epsfig,amssymb,algorithm,algpseudocode,amsthm,cite,enumerate,float,textcomp,hyperref,tikz,pgfplots}
\DeclareTextSymbol{\degre}{OT1}{23}

\usepackage{tikz}
\usetikzlibrary{calc,trees,positioning,arrows,chains,shapes.geometric,%
    decorations.pathreplacing,decorations.pathmorphing,shapes,%
    matrix,shapes.symbols}

\tikzset{
>=stealth',
  punktchain/.style={
    rectangle, 
    rounded corners, 
    draw=black, very thick,
    text width=10em, 
    minimum height=3em,
	minimum width=10em,
    text centered, 
    on chain},
circle_chain/.style={
    circle, 
     fill=black!10,
    draw=black, very thick,
    minimum size=5pt, 
    text centered, 
    on chain},
  line/.style={draw, thick, <-},
  element/.style={
    tape,
    top color=white,
    bottom color=blue!50!black!60!,
    draw=blue!40!black!90, very thick,
    text width=10em, 
    text centered, 
    on chain},
  decoration={brace},
  tuborg/.style={decorate},
  tubnode/.style={midway, right=2pt},
}

\renewcommand{\bullet}{\,\begin{picture}(-1,1)(-1,-2)\circle*{2}\hspace{0.2cm} \end{picture}\ }
	\title{{Preliminary Results on 3D Channel Modeling: From Theory to Standardization}}
\author{Abla Kammoun, Hajer Khanfir, Zwi Altman, M\'erouane Debbah, Mohamed Kamoun
\thanks{A.~Kammoun and M.~Debbah are with the Alcatel-Lucent Chair on Flexible Radio, SUPELEC, Gif-sur-Yvette, France (e-mail: \{abla.kammoun,  merouane.debbah\}@supelec.fr). H.~Khanfir and Z.~Altman are with Orange labs, France (e-mail: \{hajer.khanfir,zwi.altman\}@orange.com), M. Kamoun is with CEA-LIST, Communicating Systems Laboratory, F-91191 Gif-sur-Yvette, France (e-mail: \{mohamed.kamoun@cea.fr\})}}

\begin{document}

\maketitle
%\thanks{ This research has been supported by the ERC Starting Grant 305123 MORE (Advanced Mathematical Tools for Complex Network Engineering).}

\begin{abstract}
	Three dimensional beamforming (3D)  (also  elevation beamforming) is  now gaining a growing interest among researchers in wireless communication. The reason can be attributed to its potential to enable a variety of strategies  like sector or user specific  elevation beamforming and  cell-splitting. Since these techniques cannot be directly supported by  current LTE releases, the 3GPP is now working on defining the required technical specifications. In particular, a large effort is  currently made  to get accurate 3D channel models that support the elevation dimension. This step is necessary as it will evaluate the potential of 3D and FD(Full Dimensional) beamforming techniques to benefit from the richness of real channels. This work aims at presenting the on-going 3GPP study item {\it \bf  "Study on 3D-channel model for Elevation Beamforming and FD-MIMO studies for LTE"}, and positioning it with respect to previous standardization works. 
\end{abstract}
\section{Introduction}

Channel modeling is a fundamental step that allows performance evaluation of transmission techniques.% Considering the high cost and complexity of emerging technologies prototypes, it is no surprise that simulations have remained the most retained approach to evaluate different candidate features. To be reliable, simulations need however to use an accurate modeling of each element in the communication chain.
 While the behavior of the transmitter and the receiver are well understood, the channel, being intrinsically dependent on the surrounding environment, is much more difficult to analyse. This has triggered an increasing activity around channel modeling.

%%%%%%% Give brief overview on channel modeling in the literature %%%%%
A close look to the evolution of the theory of channel modeling reveals that this field has often been  influenced by  aspects that were important for the technology available at the time. As a consequence, first channel models used to concentrate on the large scale fading (like Okumara-Hata \cite{hata}, Lee's model \cite{lee}), whereas the small scale fading is reduced to the Doppler effect caused by receiver mobility \cite{jakes-71}. Soon after, with the emergence of wideband and multi-antenna technologies, it has been realized that small scale fading should account for the multiple reflections of the emitted signal as well as the spatial correlation between the antenna elements. 

In parallel to these theoretical works, there has been a large interest in developing within the framework of wireless standards, reference channel models which serve to define conventional ways to generate channels. For companies, these channel models are of fundamental importance since they allow them to be calibrated to the same channel conditions, and as such to assess  system-level and link-level performances of advanced signal processing techniques over real-like channels.

Like their counterparts developed in theoretical works, these channel models  have evolved in such a way to address the challenges of wireless communication technologies. 
%Moreover, standardized channel models have known a similar evolution. 
 %%%%%%%%%%%%%%%%%%%%%%%%%%%%%%%%%%%%%%%%%%%%%%%%%%%%%%%%%%%%%%%%%%%%%%%
The first standardized  models were developed in the framework of COST 207 actions \cite{Failli} which had essentially served to the standardization of GSM. Several actions had been then proposed leading to important models like the COST 231 Walfish Ikegami and COST 231 Hata channel models \cite{Damosso} but it was only with the COST 259 that the spatial structure of the radio channel was taken into account \cite{Correia}. This was a key step that paved the way towards the modeling of MIMO channels. In fact, influenced by COST actions, 3GPP Spatial Channel Model (SCM) \cite{3gpp}, Extended SCM \cite{extended-3gpp} Spatial Channel Model (SCME)  and WINNER II  \cite{winner} have emerged as low complexity alternatives to COST based channel models. While COST actions have continued through COST 273 and COST 2100 to adopt a universal involved concept, SCM and WINNER II  introduce several simplifications in order to facilitate system-level simulations. The same simplified approach has been also taken into account in IMT-advanced standardization, strongly inspired by WINNER II and 3GPP2/SCM \cite{itu}.

To further enhance  performance, the actual trend is to exploit the channel's degrees of freedom in the elevation direction. Given that most  existing channel models are only two dimensional  in that they assume that the wave propagates in the azimuth plane, a large effort in channel modeling has to be made in order to account for the impact of the channel component in the elevation direction.  This results in what we refer to as 3D beamforming or Full Dimension MIMO (FD-MIMO). One way to exploit the additional degree of freedom of 3D channels is to adapt for each user the beam pattern in the vertical direction, thereby improving the signal strength at the receiver and at the same time reducing the  interference to  other users.  Initial implementations of this technology support the potential of this technique for yielding significant gains in real indoor and outdoor deployments \cite{koppenborg}. Encouraged by these preliminary results, an extensive research activity on 3D channel modeling is being carried out by both theoretical researchers and industrials. At the time of writing this paper, the TSG-RAN-WG1 of the 3GPP is working on defining next generation channel models in the frame of the study item on elevation beamforming \cite{3gpp-study}. The outcome of this study will be used for the evaluation of advanced antenna technologies such as 3D beamforming. 

{\it Previous surveys on channel modeling and contributions:} 
Channel modeling has been the subject of many surveys. Without tackling multi-antenna modeling, \cite{rappaport} provides a good overview on propagation and large as well as small scale fading. On the other hand, \cite{molish-book} provides a comprehensive introduction to wireless channel modeling including propagation modeling and statistical description of channels. A detailed overview on propagation modeling with an exclusive summary of measurement parametrization as well as validation results has been presented in  \cite{correai}. Very recently, the authors in \cite{bruno} provide a complete coverage of recent MIMO channel propagation models as well as a description of the most recent signal processing techniques for single-cell/single-user as well as multi-user/multi-cell systems. 

Building on these  references, our objective in this work is to shed light on the current 3GPP activity around 3D beamforming and FD-MIMO, an aspect that has not been extensively covered to our knowledge. Our recent participation in 3GPP meetings  \cite{R1-133719,R1-133720} has enabled us to enhance our understanding of the current industrial challenges as well as to deeply comprehend the standards' vision for 3D beamforming.  We  think that our work contributes to cover new aspects that have not yet been investigated by a research publication. In particular, we adopt a constructive approach that starts from the basics of channel modeling to introduce in a second step standardized channels. We organize our paper as follows. We describe in section \ref{sec:channel_model} the mathematical modeling of time-varying SISO and MIMO channels. Then, we present in section \ref{sec:standards} an overview of the most known standardized channels  before discussing in section \ref{sec:3gpp} the ongoing work of the 3GPP on channel modeling.   

% In addition to drawing a comparison between the most used models in practice, we report on the progress of the 3GPP activities around the calibration of  the future 3D channel. 

%{\it Paper Organization}

%This paper is organized into two principal parts. The first one introduces briefly the basics of channel modeling, while the second part focuses on the standardization work that has been carried out so far. 
%All the described  works constitute excellent references for channel modeling, 

% Typically, each antenna port at the base station is composed of a large number of antenna elements which are vertically stacked in order to achieve a high antenna gain. 

\section{Channel models: From SISO To MIMO channel models}
\label{sec:channel_model}
Most existing channel models  account for two principal large scale effects which are:
\begin{enumerate}
\item Path loss: Path loss is an average reduction in power due to the propagation of the electromagnetic wave for a given system configuration. It increases exponentially  with distance and carrier  frequency. In general,  an approximative relation that describes the variation of the path loss with these parameters is determined through measurements. It can also include a loss caused by building penetration for indoor users \cite{3gpp,winner}. 
\item Large scale fading (shadowing): It is an average reduction in power due to the effect of shadowing caused by obstacles, and is mostly modeled as a log-normal distribution \cite{pahlavan}.  
\end{enumerate}
In the sequel, we denote by $\sigma_{SF}$ and $PL$ the incurred loss caused, respectively, by shadow fading and path loss in ${\rm dB}$. We will also focus on the downlink channel between the base station (BS) and the user equipment (UE). 
\subsection{SISO Channel Model}

{\it Non-time dispersive channels}

Non-time dispersive channels do not spread the channel in time. The channel can be viewed as a distortion that scales the emitted wave. Its impulse response at time $t$ to a unit impulse $\delta(.)$ transmitted at time $t-\tau$ is given by: 
%In case of Narrowband transmissions,  It is given by:
%Assume that the emitted electromagnetic wave can be modeled as a plane wave with an electric at field at time $t$ and direction $\vec{\bf r}(t)$ given by:
%$$
%\vec{E}(t)=\vec{E}_0(t) \exp(\jmath (2\pi f_c t- \vec{\bf k}\bullet \vec{\bf r}(t))
%$$
%where $f_c$ denotes the carrier frequency and $\vec{\bf k}$ the wave vector. In narrowband channel modeling, only the first path is accounted for. If we suppose that the propagation delay is too small, the received electromagnetic wave is then given up to a constant phase by:
%$$
%\vec{E}_r(t)=10^{-\frac{SF+PL}{10}}\vec{E}_0(t) \exp(\jmath (2\pi f_c t+ \vec{\bf k}\bullet \vec{\bf r}_{UE}(t))
%$$
%where $\vec{\bf r}_{UE}(t)$ denotes the position of the mobile station at time $t$ in the coordinate system corresponding to the UE. (the sign (-) becomes (+) since the wave vector in the coordinate system at the receiver is the opposite of that at the transmitter). 
%The narrowband channel which stands for the distortion undergone by the emitted wave is thus given by:
\begin{equation}
	h(t,\tau)=10^{-\frac{\sigma_{SF}+PL}{10}}\delta(\tau-\tau_0)\exp( \jmath\vec{\bf k}\bullet \vec{\bf r}_{UE}(t))
\end{equation}
where $\vec{\bf r}_{UE}(t)$ denotes the position of the mobile station at time $t$ in the coordinate system corresponding to the UE and $\vec{\bf k}$ is the wave vector such that $\|{\bf k}\|=\frac{2\pi f}{c}$ where $c$ is  the speed of light, and $\tau_0$ is the channel delay.
If the UE is moving with a constant velocity $\vec{\bf v}_0$ then $\vec{\bf r}_{UE}(t)=\vec{\bf v}_0 t+\vec{\bf r}_{UE,0}$ ($\vec{\bf r}_{UE,0}$ being the position of the UE at time $0$), the resulting channel is  given by:
\begin{equation}
	h(t,\tau)=10^{-\frac{\sigma_{SF}+PL}{10}}\exp(\jmath \vec{\bf k}\bullet \vec{\bf v}_0 t)\exp(\jmath \vec{\bf k}\bullet \vec{\bf r}_{\rm UE,0}).
\label{eq:doppler}
\end{equation}
Equation \eqref{eq:doppler} reveals that because of the user mobility, the effective frequency is shifted by a $f_D= \frac{\vec{k}\bullet{\vec{\bf v}_0}}{c}$. This phenomenon is referred to as the Doppler effect.

{\it Time-Dispersive channel models}

In time-dispersive channel models, the propagation delay cannot be neglected  compared to the signal period. The received wave is therefore the sum of the contribution stemming from several paths. Each path corresponds to a single or several reflections of the incident wave, \cite{petrus02,chen03}. In this case the channel impulse response is obtained by summing the contribution of  all possible paths. 
\begin{equation}
	h(t,\tau)=10^{-\frac{\sigma_{SF}+PL}{10}}\sum_{l=1}^{N_C} \alpha_l(t,\tau_l) \delta(t-\tau_l).
\label{eq:multi-path}
\end{equation}
where $\alpha_l(t,\tau_l)$ and $\tau_l$ represent respectively the additional attenuation and the delay corresponding to the $l$-th path. Note that in this case, multipath propagation causes small-scale fading. The received power can suffer severe fading, and additional amount of power, known as fade margin, should be provided in order to enhance the system quality. These aspects are beyond the scope of this paper but the interested reader can refer to \cite{cardoso-1} and \cite{cardoso-2} for further details.
%Assuming that the wavelength is much smaller than obstacles, propagation of electromagnetic waves can be studied using geometrical optics. The received wave is therefore the sum of the contribution stemming from several paths. Each path corresponds to a single or several reflections of the incident wave, referring respectively to single-bounce \cite{petrus02} or multiple-bounce based models \cite{chen03} (See fig. \ref{fig:single} and fig. \ref{fig:multiple}).
\subsection{MIMO channel Models}
As we have noted above,  SISO channel models are fully characterized by the statistics of the delays, powers and phases  corresponding to each path.  However, this  information  becomes highly insufficient to capture the richness of MIMO channels. 
\subsubsection{Double Directional MIMO Channel}

To describe MIMO channels, it can be useful to distinguish between the radio channel and the propagation channel which only depends on the environment and excludes as such the effect of antenna responses.  This propagation channel is referred to as double-directional channel model, \cite{steinbauer,molisch-04}. It does not depend on the number of antennas at the receiver or the transmitter. Actually, it is  a scalar or a $2\times 2$ matrix if dual-polarization is considered.  

In addition to the delays, the double directional model depends on the directions of departure and the directions of arrival. The double directional impulse function corresponding to the $\ell$-th multipath component (MPC) is thus given by:
\begin{equation}
h_\ell(t,\tau,\Omega,\Psi)=\alpha_\ell\delta(\tau-\tau_\ell) \delta(\Omega-\Omega_l)\delta(\Psi-\Psi_l)\exp(\jmath \vec{\bf k}_{\ell,r}\bullet \vec{\bf r})
\end{equation}
where $\tau$ is the delay variable and $\Omega$, $\Psi$ stand for the spatial angles respectively at the transmitter and  the receiver, i.e., $\Omega=\left(\phi,\theta\right)$ and $\Psi=\left(\varphi,\vartheta\right)$ where $\phi$ and $\varphi$ are the departure and arrival azimuth angles whereas $\theta$ and $\vartheta$ are the departure and arrival elevation angles (See Fig. \ref{fig:3gpp_3d}). 

Besides, $\alpha_l$ is the complex amplitude of the path $l$ whereas $\vec{\bf r}$ is the position of the receiver at time $t$ and $\vec{\bf k}_{\ell,r}$ is the wave vector corresponding to the received wave vector of the $\ell$ th path. 

\begin{figure}[t]
	\begin{center}
	\includegraphics[width=0.8\columnwidth]{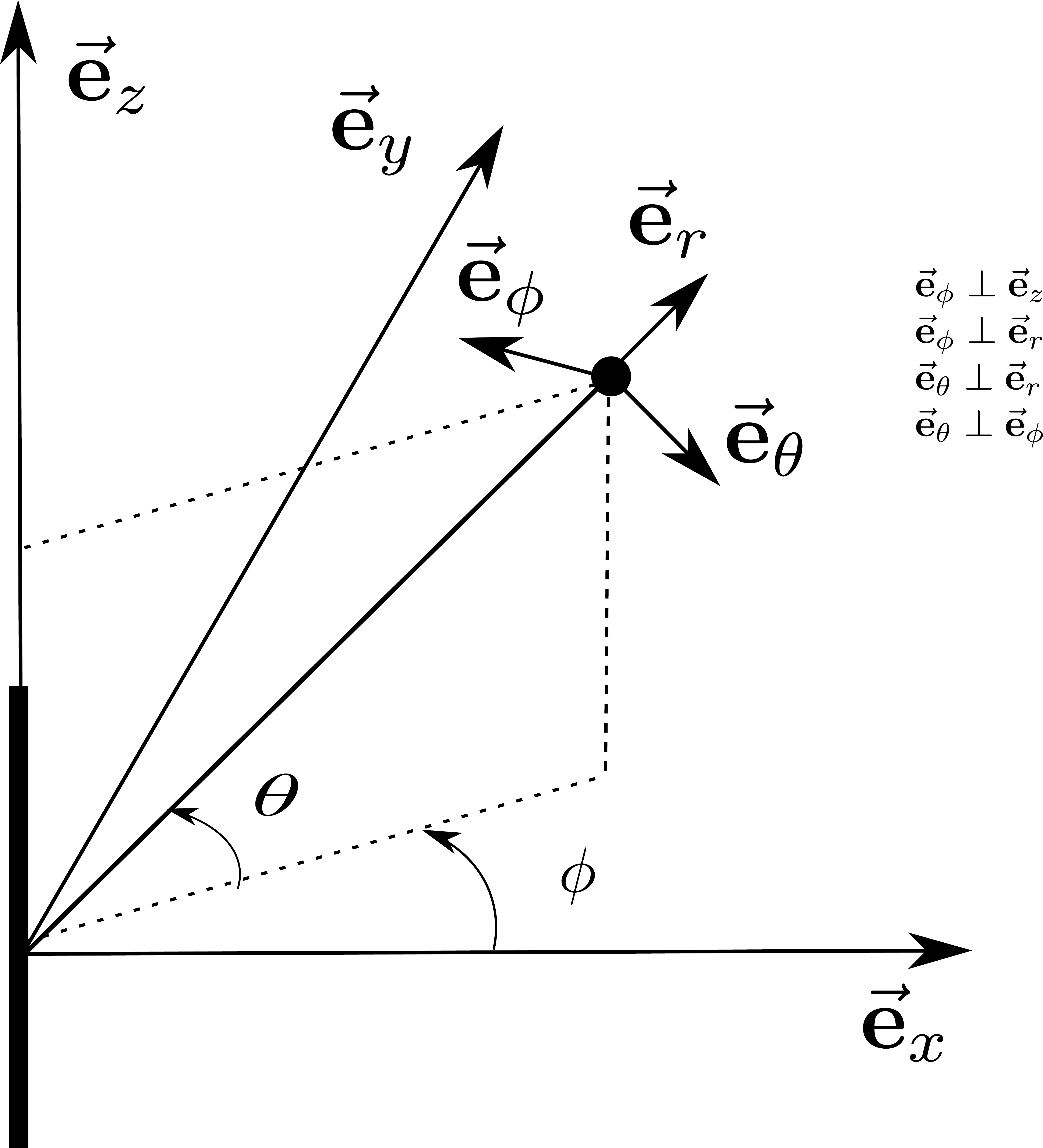}
	\caption{A vertically polarized antenna element}
	\label{fig:system_spherique}
\end{center}
\end{figure}

The double directional impulse response is  the sum of the $N_C$ MPCs and is given by:
\begin{equation}
	h{(t,\tau,\Omega,\Psi)}=\sum_{l=1}^{N_C}\alpha_\ell\delta(\tau-\tau_\ell) \delta(\Omega-\Omega_l)\delta(\Psi-\Psi_l)\exp(\jmath \vec{\bf k}_l\bullet \vec{\bf r})
\end{equation}
If polarization is taken into account, $\alpha_\ell$ and thus $h{(t,\tau,\Omega,\Psi)}$ are $2\times 2$ matrices which describe the coupling between vertical and horizontal polarizations \cite{molish}: 
\begin{equation}
{\bf h}(t,\tau,\Omega,\Psi)=\begin{bmatrix}
	h^{VV}(t,\tau,\Omega,\Psi) &   h^{VH}(t,\tau,\Omega,\Psi) \\
	h^{HV}(t,\tau,\Omega,\Psi) &  h^{HH}(t,\tau,\Omega,\Psi). 
\end{bmatrix}
\end{equation}
In other words, the elements  $h^{VV}(t,\tau,\Omega,\Psi) $ and $h^{VH}(t,\tau,\Omega,\Psi)$ represent up to a scalar  what would be obtained by a  receiver in the vertical and horizontal directions if the transmitted wave is vertically polarized.
In the same way, the elements $h^{HH}(t,\tau,\Omega,\Psi) $ and $h^{HV}(t,\tau,\Omega,\Psi)$ represent up to scalar what would be received in the horizontal and vertical polarizations if the transmitted wave is horizontally polarized. For the reader convenience, we recall that  If $(\vec{\bf e}_r,\vec{\bf e}_\theta,\vec{\bf e}_\phi)$ is the spherical coordinate system associated to the vertical to the ground $\vec{\bf e}_z$, vertical polarization refers to the polarization along $\vec{\bf e}_\theta$ whereas horizontal polarization refers to the polarization along  $\vec{\bf e}_\phi$. In particular, a vertically oriented antenna elements delivers an electric field which is along $\vec{\bf e}_\theta$ (See Fig. \ref{fig:system_spherique}).

%In case of polarized channels, $h_\ell(t,\tau,\Omega,\Psi)$ becomes a $2\times 2$ matrix given by:
%$$
%h_\ell(t,\tau,\Omega,\Psi)={\bf a}_\ell\delta(\tau-\tau_\ell) \delta(\Omega-\Omega_l)\delta(\Psi-\Psi_l)\exp(\jmath \vec{\bf k}_l\bullet \vec{\bf r})
%$$
%where ${}
 %Most channel standards assume 
%\begin{center}
%\begin{figure}[t]
%\begin{center}
%	\begin{minipage}[b]{\textwidth}
%\includegraphics[width=0.45\columnwidth]{single_bounce.pdf}
%		\caption{Single bounced channel models}
%		\label{fig:single}
%\end{minipage}
%	\begin{minipage}[b]{\textwidth}
%\includegraphics[width=0.45\columnwidth]{multiple_bounce.pdf}
%		\caption{Multiple bounced channel models}
%		\label{fig:multiple}
%	\end{minipage}
%\end{center}
%\end{figure}
%\end{center}
%

\subsubsection{Radio channel}
\begin{figure}[t]
	\includegraphics[width=\columnwidth]{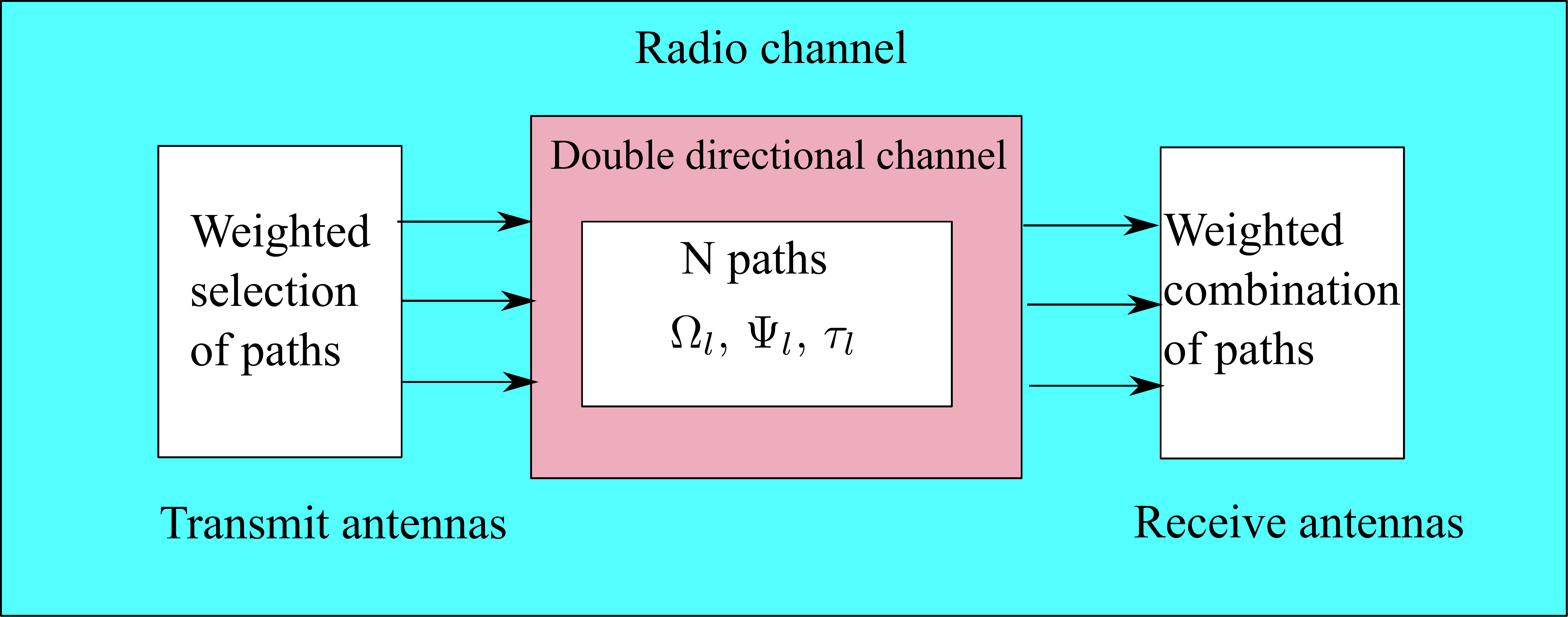}
	\caption{The radio channel}
	\label{fig:radio_channel}
\end{figure}

The radio channel is obtained by incorporating the effect of the antennas. This can be modeled at the reception or the transmission side as a coherent sum over all directions, (see Fig.\ref{fig:radio_channel}). 
Let $N_{T}$ and $N_{R}$ denote the number of the transmitting and receiving antennas. The radio channel is a $N_R\times N_T$ matrix given by:
\begin{equation}
{\bf H}(t,\tau)=\int  \vec{\bf g}_r(\Psi)^{\mbox{\tiny T}} h(t,\tau, \Omega,\Psi) \vec{\bf g}_T(\Omega) \vec{\bf a}_R(\Psi)\left(\vec{\bf a}_T(\Omega)\right)^{\mbox{\tiny T}} d\Omega d\Psi
\end{equation}
where:
\begin{itemize}
	\item $\vec{\bf g}_r$ and $\vec{\bf g}_T$ are the patterns of the receiving and transmitting antennas. When polarization is considered, $\vec{\bf g}_r(\Psi)$ and $\vec{\bf g}_T$ are $2\times 1$ vectors whose entries represent the vertical and horizontal field patterns.  
	\item Vectors $\vec{\bf a}_R(\Psi)$ and $\vec{\bf a}_T(\Omega)$ are the array responses of the transmitting and receiving antennas whose entries are given by:
		\begin{align}
			\left[\vec{\bf a}_R(\Psi)\right]_i&=\exp\left(\jmath \vec{\bf k}_{R,\Psi}\bullet \vec{\bf x}_{R,i}\right) \\
			\left[\vec{\bf a}_T(\Omega)\right]_i&= \exp\left(\jmath \vec{\bf k}_{T,\Omega}\bullet \vec{\bf x}_{T,i}\right)
	\end{align}
	\end{itemize}
	where $\vec{\bf x}_{R,i}$ is the location vector   of the $i$th receiving antenna whereas $\vec{\bf x}_{T,i}$ is that of the $i$th transmitting antenna. These location vectors are computed with respect to the global cartesian coordinate system $(\vec{\bf e}_x,\vec{\bf e}_y,\vec{\bf e}_z)$, (See Fig. \ref{fig:system_spherique}).
	Substituting $h(t,\tau,\Omega,\Psi)$ by its expression, the radio channel writes as:
	\begin{equation}
		{\bf H}(t,\tau)=\sum_{\ell=1}^{N_C}  \delta(\tau-\tau_\ell)\vec{\bf g}_r(\Psi_\ell)^{\mbox{\tiny T}} \boldsymbol{\alpha}_l \vec{\bf g}_T(\Omega_\ell) \vec{\bf a}_R(\Psi_\ell)\left(\vec{\bf a}_T(\Omega_\ell)\right)^{\mbox{\tiny T}}
	\label{eq:radio_channel}
\end{equation}
Equation \eqref{eq:radio_channel} is generic in that it encompasses several kind of channel models. In particular, if the spatial angle $\Omega$ or $\Psi$ are defined using only the azimuth, the channel is said to be two-dimensional (2D), whereas (3D) channels are obtained by accounting for both elevation and azimuth angles. 
%Equation \eqref{}, corresponds to the channel between physical antenna elements.

\section{Standardized channel models}
\label{sec:standards}
\subsection{Generation of the channel  in system level approach based standards}
\begin{figure}
	\includegraphics[width=\columnwidth]{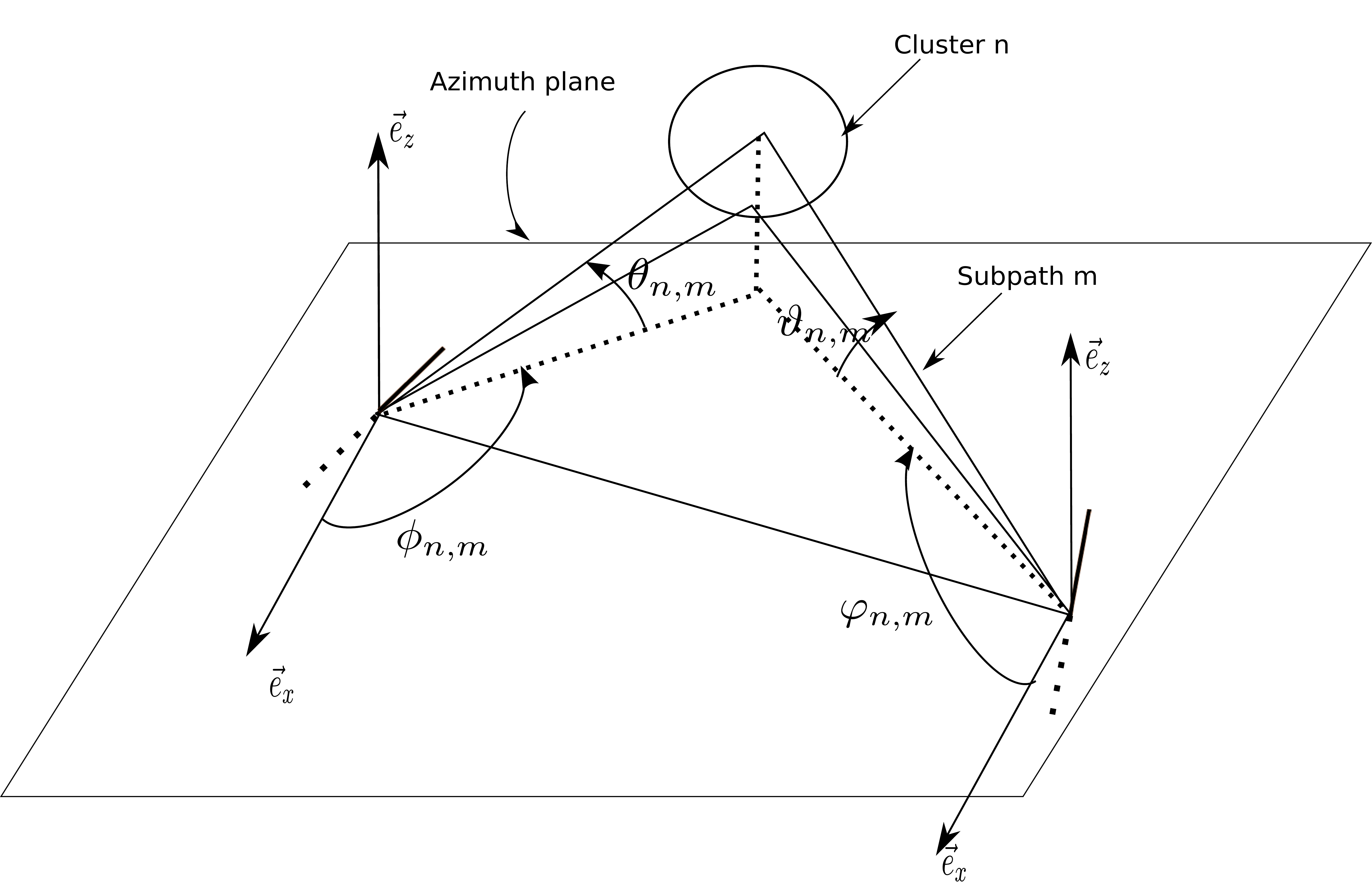}
	\caption{3D channel model}
	\label{fig:3gpp_3d}
\end{figure}

Building on the theoretical framework of channel modeling, standards based on a system level approach rely on the same findings described by \eqref{eq:radio_channel}. In a similar way as above, the channel is composed of many propagation paths with different time delays, referred to as  {\it clusters}. Each cluster is characterized by three quantities: the delay $\tau_n$, the spatial angle of departure $\Omega_n=\left(\theta_n,\phi_n\right)$ and the spatial angle of arrival $\Psi_n=\left(\vartheta_n,\varphi_n\right)$. However, the standardized channel models differ from those described in the previous section in that it is assumed that each cluster gives rise to $M_n$ unresolvable paths which have the same delay as the original cluster. Moreover, each sub-path is characterized by its spatial angles $\Omega_{n,m}=\left(\theta_{n,m},\phi_{n,m}\right)$ and $\Psi_{n,m}=\left(\vartheta_{n,m},\varphi_{n,m}\right)$ where 
\begin{align}
	\theta_{n,m} &=\theta_n+ c_\theta \alpha_m , &	\phi_{n,m} = \phi_n +c_\phi \alpha_m \nonumber \\
	\vartheta_{n,m}&=\vartheta_n +c_\vartheta \alpha_m, &\varphi_{n,m}=\varphi_n + c_\varphi \alpha_m
\end{align}
and $\alpha_m, m=1,\cdots,M_n$ is a set of symmetric fixed values and $c_\phi$,  $c_\theta$, $c_\varphi$ and $c_\varphi$ controls the spread inside the cluster $n$, (See Fig. \ref{fig:3gpp_3d}).
%The most important difference with the channel models described above is that we assume that each cluster give rise to $M_n$ unresolvable paths which have the same delay as the original cluster. 

The channel response corresponding to the $n$-th path is thus given by:
\begin{align}
	{\bf H}_{n}(t) &= \sqrt{10^{-(PL+ \sigma_{SF})/10}}\sum_{m=1}^{M_n} \sqrt{P_{n,m}}\vec{\bf g}_R(\varphi_{n,m},\vartheta_{n,m})^{\mbox{\tiny T}} \boldsymbol{\alpha}_{n,m}\nonumber\\
			      &\times\vec{\bf g}_T(\phi_{n,m},\theta_{n,m})  \vec{\bf a }_R\left(\varphi_{n,m},\vartheta_{n,m}\right)\nonumber\\
		&\times\vec{\bf a}_T(\phi_{n,m},\theta_{n,m})^{\mbox{\tiny T}}\exp\left(\jmath \vec{\bf k}_{r,n,m}\bullet \vec{\bf v}t\right)
\end{align}
where $\vec{\bf k}_{r,n,m}$ is the wave vector corresponding to the $m$-th subpath, $P_{n,m}$ is the power of the $m$-th subpath and 
\begin{equation}\boldsymbol{\alpha}_{n,m}=\begin{bmatrix}\exp\left(\jmath \Phi_{n,m}^{VV}\right) & \sqrt{\kappa_{n,m}}\exp\left(\jmath \Phi_{n,m}^{VH}\right) \\
\sqrt{\kappa_{n,m}}\exp\left(\jmath\Phi_{n,m}^{HV}\right)& \exp\left(\jmath \Phi_{n,m}^{HH}\right)
\end{bmatrix},
\end{equation}
$\kappa_{n,m}$ being the cross-polarization ratio and $\Phi_{n,m}^{HH}$, $\Phi_{n,m}^{HV}$, $\Phi_{n,m}^{VH}$, and $\Phi_{n,m}^{VV}$ being random phases.
If the UE is on Line of sight (LOS) with respect to the cell, it is assumed that the first cluster contains the line-of-sight (LOS) contribution of the channel. In this case, the channel response  is given by \eqref{eq:los} (equation in the top of the next page),
\begin{figure*}
	\begin{align}
		{\bf H}_{n}(t) &=\sqrt{10^{-\frac{PL+\sigma_{SF}}{10} }}\left(\sqrt{\frac{1}{K+1}}\sum_{m=1}^{M_n}\sqrt{P_{n,m}} \vec{\bf g}_R\left(\varphi_{n,m},\vartheta_{n,m}\right)^{\mbox{\tiny T}} \boldsymbol{\alpha}_{n,m} \vec{\bf g}_T\left(\phi_{n,m},\theta_{n,m}\right)\vec{\bf a }_R\left(\varphi_{n,m},\vartheta_{n,m}\right)\vec{\bf a}_T\left(\phi_{n,m},\theta_{n,m}\right)^{\mbox{\tiny T}}\right.\nonumber \\
			       &\left.\times \exp\left(\jmath t \vec{\bf k}_{r,n,m}\bullet \vec{\bf v}\right)+\delta(n-1)\sqrt{\frac{K}{K+1}}\vec{\bf g}_R\left(\varphi_{LOS},\vartheta_{LOS}\right)^{\mbox{\tiny T}} \boldsymbol{\alpha}_{LOS} \vec{\bf g}_T\left(\phi_{LOS},\theta_{LOS}\right)\vec{\bf a }_R\left(\phi_{LOS},\theta_{LOS}\right)\vec{\bf a }_T\left(\phi_{LOS},\theta_{LOS}\right)^{\mbox{\tiny T}}\right.\nonumber\\
&\times\left.\exp\left(\jmath t    \vec{\bf k}_{r,LOS}\bullet \vec{\bf v}\right)\right)\label{eq:los}
	\end{align}
\end{figure*}
where   $$\boldsymbol{\alpha_{LOS}}=\begin{bmatrix}
	\exp\left(\jmath \Phi_{LOS}^{VV}\right) &  {0} \\
	0  & \exp\left(\jmath \Phi_{LOS}^{HH}\right)
\end{bmatrix},$$ $\vec{\bf k}_{r,LOS}$ is the wave vector along the line-of-sight direction and $K$ is the rice factor \footnote{Note that \eqref{eq:los} encompasses the non-line-of-sight(NLOS) case by taking $K=0$.}. 

Most standardized channels like SCM \cite{3gpp}, SCME \cite{extended-3gpp}, WINNER \cite{winner} and ITU \cite{itu} are based on the model described by \eqref{eq:los}. It is thus no surprise that these models share almost the same procedure for generating the channel. Unlike the  COST family of  channel models, WINNER , ITU , SCM  and SCME follow a system level approach, in which without being physically positioned, clusters are described through statistical parameters, known as {\it large scale parameters}. These latter are random correlated variables drawn from given distributions, and are specific for each user. They serve to generate powers, delays and angles of each path,  which are often  referred to as {\it small scale parameters} in that they describe the channel at a microscopic level. 

\begin{figure}[h]
	\begin{center}
	\begin{tikzpicture}
  [scale=0.5,node distance=.8cm,
  start chain=going below,]
     \node[punktchain, join=by {->}] (Scenario) {Choose of scenario};
     \node[punktchain, join=by {->}] (probf)      {Random dropping of UEs};
     \node[punktchain, join=by {->}] (foreach)      {For each UE};
%     \node[punktchain, join] (investeringer)      {Investeringsteori};
%     \node[punktchain, join] (perfekt) {Det perfekte kapitalmarked};
%     \node[punktchain, join, ] (emperi) {Emperi};
%      \node (asym) [punktchain ]  {Asymmetrisk information};
 		\node[circle_chain,below=of foreach,inner sep=0pt, minimum size=2pt,join](pt1){};
        \node[punktchain, join=by {->},right=of pt1]
            (large_scale) {Generate Large scale parameters};
 		\node[circle_chain,below=of pt1,inner sep=0pt, minimum size=0pt,join=with pt1, below= of pt1](pt2){};
 		\node[circle_chain,below=of pt1,inner sep=0pt, minimum size=2pt,join=with pt2, below= of pt2](pt3){};
		\node[punktchain, right =of pt3,join=with pt3]
            (small_scale) {Generate Small Scale parameters(Path Powers and angles)};
	\node[circle_chain,below=of pt1,inner sep=0pt, minimum size=0pt,join=with pt3, below= of pt3](pt4){};
	\node[circle_chain,below=of pt1,inner sep=0pt, minimum size=2pt,join=with pt4, below= of pt4](pt5){};
	\node[circle_chain,below=of pt1,inner sep=0pt, minimum size=0pt,join=with pt5, below= of pt5](pt6){};
	\node[circle_chain,below=of pt1,inner sep=0pt, minimum size=2pt,join=with pt6, below= of pt6](pt7){};

		\node[punktchain, right =of pt5,join=with pt5]
		 (antenna) {Compute the antenna gains at each set of azimuth and elevation angles };

		\node[punktchain, right =of pt7,join=with pt7]
(generate) {Generate the channel using \eqref{eq:los} };
      \end{tikzpicture}
  \end{center}
	  \caption{Required steps for the Generation  of standards' channel models}
	  \label{fig:steps}
  \end{figure}
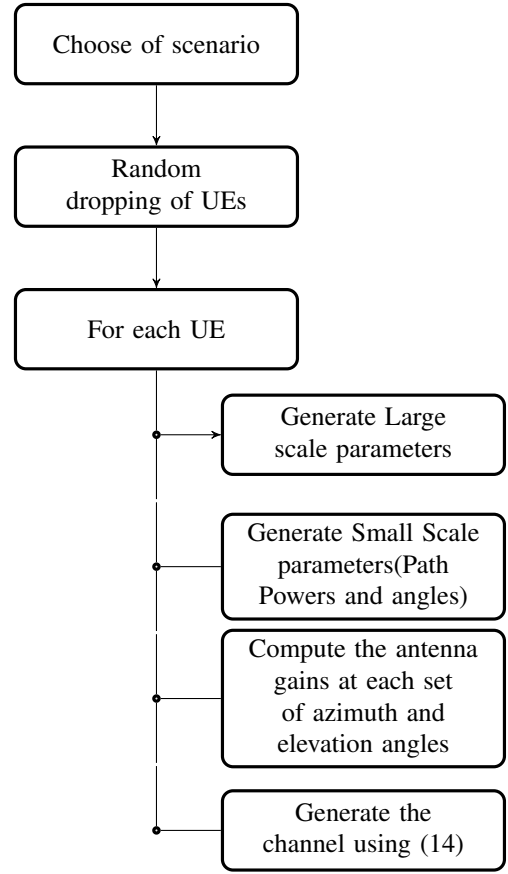

The generation of the channel in system level approach based standards  follows the following steps illustrated in fig. \ref{fig:steps}
%\begin{enumerate}
%	\item Choice of the scenario and dropping of  the UEs. For each UE,  general parameters (position, pathloss, Large scale parameters, Propagation Mode, ...) are then set. 
%	\item  Generation of the small scale parameters is performed based on the obtained large scale parameters in step 1).
%	\item With the small scale parameters on hand, the channel is computed using \eqref{eq:los}.
%	\end{enumerate}

\begin{figure}
	\includegraphics[width=\columnwidth]{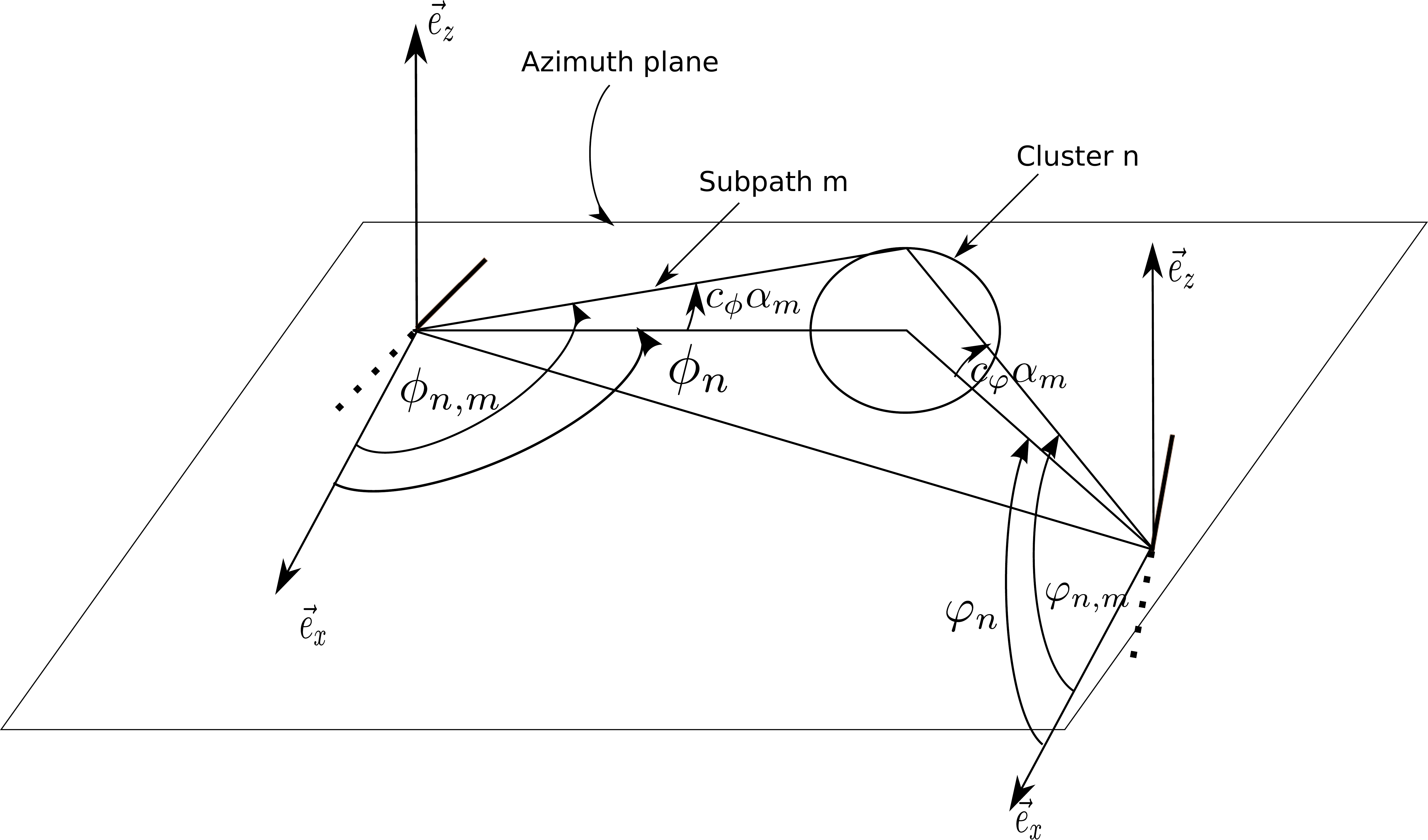}
	\caption{2D channel model}
	\label{fig:3gpp}
\end{figure}

	While system level approach based channels follow the same philosophy, they differentiate in the way the slow fading and the small scale parameters are generated. The reason is that these channels support different frequencies and are tuned using different measurement campaigns. Hereafter, we illustrate the main principal differences between these models. 
\begin{itemize}
	\item Number of scenarios:  SCM was dedicated originally to outdoor channel models. It defines three scenarios which are urban micro, urban macro and suburban macro. SCME keeps the same number of scenarios but WINNER II and ITU extends the channel to more scenarios (5 channels for ITU and 17 scenarios for WINNER II).
	\item Frequency range: Channel models depend on the frequency through the pathloss. While the SCM channel model is adjusted for frequencies of 1.9GHz, SCME, ITU and WINNER II, support a range of frequency of 2-6 Ghz. 
	\item Bandwidth and number of clusters: The SCM channel was targeted to up to 5MHz RF bandwidth. Thus, only $6$ clusters with different delays were found to be sufficient. SCME which is an extension of SCM to bandwidths up to $100$Mhz modifies the number of effective clusters by subdividing each cluster to 3 or 4 sub-clusters with different delays, while keeping the same number of multi-path components. In WINNER II and ITU, the number of clusters is increased and depends on the scenario, whereas the two strongest clusters are subdivided using almost the same technique proposed in SCME.   
	\item Propagation mode: In SCM, the LOS propagation mode is defined only in the ubran micro scenario, whereas only the NLOS is considered for the other scenarios. On the other hand, WINNER II defines for some specific scenarios  a third propagation mode namely the obstructed-line-of-sight (OLOS) to describe the situation where the  LOS direction is obstructed but still remains dominant. Moreover, it   usually supports both propagation mode (LOS and NLOS) except for some scenarios where only the LOS or the NLOS is allowed.  On the contrary, ITU always supports LOS and NLOS propagation modes for its five scenarios. It also defines a third propagation mode in the UMI scenario namely the outdoor to indoor propagation mode which serves to describe the situation where the UE is inside a building. This is to be compared with WINNER II which defines a fully-fledged scenario \cite{winner} (scenario B4) to account for the outdoor to indoor link.
	\item 2D vs 3D channel models: Almost all the system level approach based standards are 2 dimensional. This implies that all the waves are in the azimuth plane (plane parallel to the ground). The unique 3D standard based on the system level approach is WINNER+  \cite{D5.3} which is an extension of WINNER II to the three dimensional case. The difference between three and two dimensional models is illustrated in Fig. \ref{fig:3gpp_3d} and Fig. \ref{fig:3gpp}.
	\item Large scale parameters: Large scale parameters describe the channel at a macroscopic level in that they control the distributions of small scale parameters. They are random correlated variables that are generated for each link between a UE and a site (See \eqref{sec:system}). In 3GPP SCM, three large scale parameters are considered which are the delay spread, the departure azimuth angular spread and the shadow fading. To these large scale parameters, WINNER II and ITU  add the arrival azimuth spreads and the rician factor, thereby increasing the number of large scale parameters to $5$. By considering elevation, WINNER+ accounts for $7$ large scale parameters which include the departure and arrival elevation spreads. 
	\end{itemize}
%For sake of illustrate, let us consider the large scale parameters in SCM. SCM defines three large scale parameters which are the delay spread $\sigma_{DS}$, the azimuth angular spread $\sigma_{AS}$ of departure and the shadow fading $\sigma_{SF}$.

\subsection{Antenna configuration}

\begin{figure}[t]
	\begin{center}
	\includegraphics[width=0.5\columnwidth]{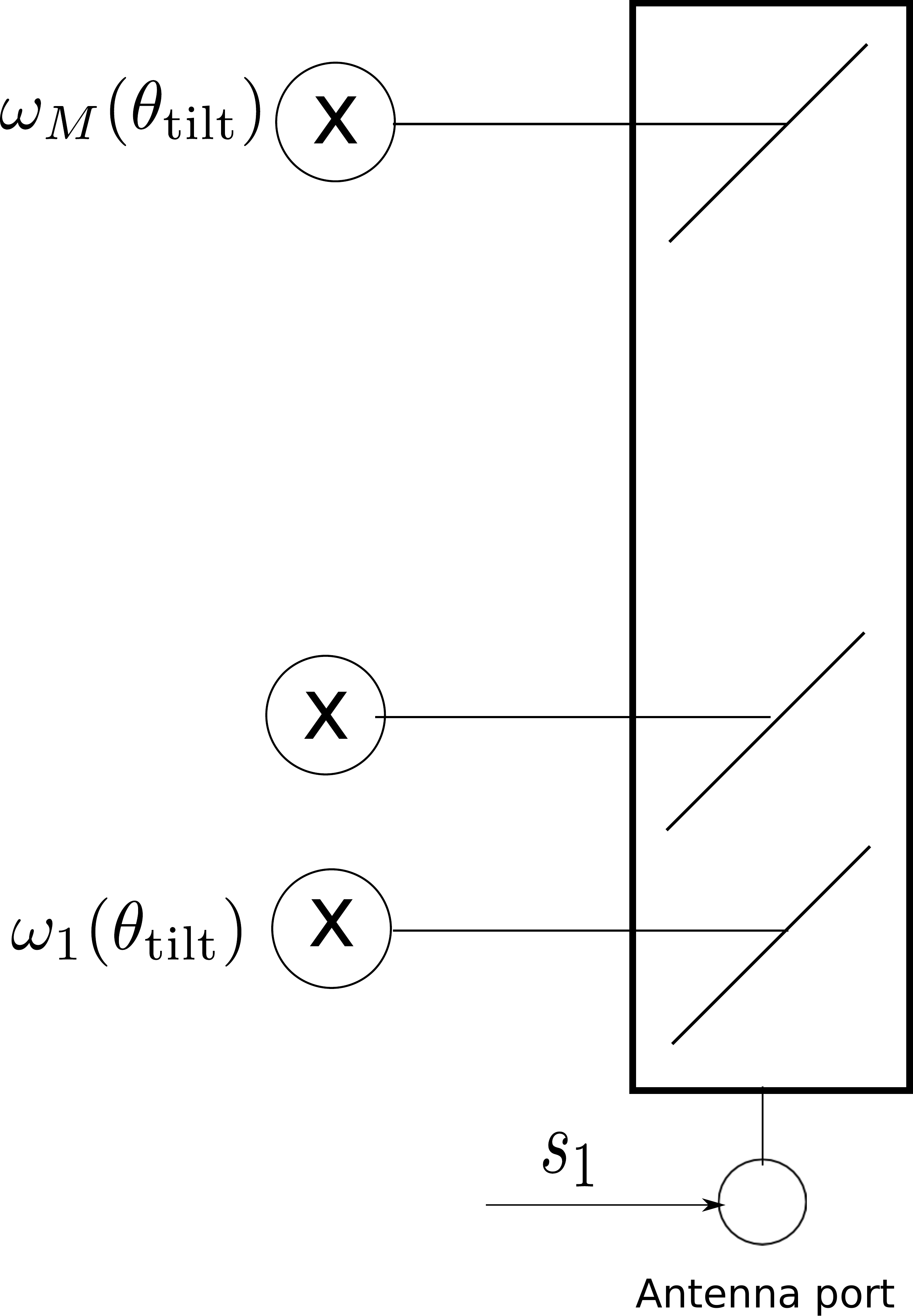}
	\caption{Antenna port}
	\label{fig:antenna_port}
\end{center}
\end{figure}

To make it easier for the reader to understand the use of the 3D beamforming technique in practice, we shall introduce in this section the antenna configuration that is being proposed in current standardization works. %{\it Antenna port, Antenna element}
In LTE, the radio resource is organized on the basis of antenna ports. Each antenna port is mapped to a group of physical antenna elements which carry the same signal. As a consequence, each antenna port is viewed as a single antenna at the receiver side. In LTE, these antenna ports are used to support different transmission modes \cite{burbank}. %The use of beamforming technique corresponds to modes 7 and 8.%where respectively ports 5 (mode 7) and ports 7 and 8 (mode 8) are involved, \cite{burbank}. 
For instance, in transmission mode 7 (TM7), antenna port 5 is used to transmit only one stream, a configuration which is often referred to as single layer transmission.  This antenna port is mapped to more than one antenna element. The signal is fed to each of the antenna elements with a corresponding weight in order to focus the wavefront in the direction of the UE. Since  the weights are not extracted from a code-book, this technique is called non-codebook based precoding (See fig. \ref{fig:antenna_port}).
At the receiver side, each antenna port appears as a single antenna, because its elements carry the same signal. It is thus no surprise that we are interested rather on the channel between the transmitting antenna port and the receiver side. For that, in \eqref{eq:los} it is the antenna port pattern that should be used and not the antenna element pattern. 

In theory, the antenna port pattern depends on the number of antenna elements, their patterns, as well as their relative positions and their corresponding weights.  To enable an abstraction of the role played by the antenna elements to perform the downtilt, ITU channel model approximates the pattern of each antenna port by a narrow beam in the elevation plane with $\theta_{\rm 3dB} = 15\degre$.
% because the channel is only a function of the tilt angle and not weights. 
%To use \eqref{eq:los}, the retained approach in practice is to approximate the pattern of each antenna port. 
%The use of 3D antenna patterns  
%Since setting fixed tilt angles was the most common approach, standards used to approximate the pattern of the transmitting antenna port by using a combined pattern which discards the effect of side lobes.
% In ITU for example, the pattern at the transmitting antenna port is approximated by a narrow beam in the elevation plane.
In fact the combined pattern considered in ITU is given by:
%{\it Antenna pattern } Unlike the 3GPP SCM  which is based on 2D antenna patterns, ITU channel standard provides a 3D pattern for each antenna port. After discarding the effect of side lobes, the combined radiation pattern at the output of the antenna port writes as:
\begin{align}
	A_{P}\left(\phi,\theta\right)&=G_{P,{\rm max}}-\min\left\{-\left(A_H(\phi)+A_V(\theta)\right),A_{ m}\right\} ,\nonumber\\
																			  &  A_{m} =20 {\rm dB} ,  G_{P,{\rm max}}=17 {\rm dBi}
\end{align}
where
\begin{align}
	A_H(\phi)&=-\min \left[12\left(\frac{\phi}{\phi_{\rm 3dB}}\right)^2,A_m\right]\nonumber \\ 
	&\phi_{\rm 3dB}= 70\degre 
\end{align}
and 
\begin{align}
	A_V(\theta)&=-\min \left[12\left(\frac{\theta-\theta_{\rm tilt}}{\theta_{\rm 3dB}}\right)^2,A_m\right] \nonumber \\
	&  \theta_{\rm 3dB} =15 \degre
\end{align}
where ${\theta}_{\rm tilt}$ is the electrical tilt angle. %The channel between the receiving antenna and each transmitting antenna port can be described by using \eqref{eq:los}. 
This approach despite its simplicity poses a problem when dual-polarized channels are used. In fact, while standards provide the expression of the global pattern at each antenna port, they do not show how one can deduce the horizontal and vertical field patterns. To circumvent this difficulty, the document 3GPP 36.814 \cite[section A.2.1.6.1, page 79]{36814} proposes to model the polarization {\it as angle independent  in both azimuth and elevation}. Recall that in the case of dual-polarized channels, each antenna port is composed of cross-polarized antenna elements that  are slanted in the plane perpendicular to the boresight (direction of the maximum gain) by a slant angle $\alpha$.

If $A(\phi,\theta)$ denotes the global pattern of the transmitting antenna port when each element is vertically polarized, then  after slanting each antenna element by a slant angle $\alpha$ in the plane perpendicular to the boresight, the global pattern along horizontal and vertical polarizations is approximated according to the document 3GPP 36.814 by:
\begin{equation}
{\bf g}_T(\phi,\theta)=\left[\sqrt{A(\phi,\theta)_{\rm lin}} \cos\alpha, \sqrt{A(\phi,\theta)_{\rm lin}} \sin\alpha\right]
\end{equation}
Obviously, such an approximation does not hold in practice. It is only valid at the boresight direction. To determine the right decomposition of the antenna pattern on the vertical and horizontal polarizations, one has to know exactly the exact structure of the antenna port, i.e, the exact number of antenna elements, as well as their relative positions. This information was missing, because the adopted strategy was to rely on the channel with each transmitting antenna port  rather than that with each transmitting antenna element, \cite{37840}. 

%Hopefully, this policy is being changed to the more flexible one which is build on the pattern of the antenna element. 

%\subsection{Physical channel model with elevation beamforming}
%{\it Beamforming}
\section{ Future 3GPP channel model}
\label{sec:3gpp}
\subsection{3D beamforming}
At the time of the writing of this paper, the TSG-RAN-WG1 is  investigating the use of 3D beamforming for next generation systems. In order to enable  optimization of this technique, the channel between antenna elements rather than that between antenna ports is considered. It is also assumed that antennas are arranged in a 2D array where each column contains $M$ antenna elements.  There are exactly $K$ antenna elements per antenna port with a pattern $A_E$ given by:
\begin{align}
	A_{E}\left(\phi,\theta\right)&=G_{E,{\rm max}}-\min\left\{-\left(A_H(\phi)+A_V(\theta)\right),A_{ m}\right\} ,\nonumber\\
																			  &  A_{m} =30 {\rm dB} ,  G_{E,{\rm max}}=8 {\rm dBi}
\end{align}
where
\begin{align}
	A_H(\phi)&=-\min \left[12\left(\frac{\phi}{\phi_{\rm 3dB}}\right)^2,A_m\right] \nonumber\\ 
				   &\phi_{\rm 3dB}= 65 ^{\circ} 
\end{align}
and 
\begin{align}
	A_V(\theta)&=-\min \left[12\left(\frac{\theta-\theta_{\rm tilt}}{\theta_{\rm 3dB}}\right)^2,SLAV\right] \nonumber \\
					 &  SLAV = 30, \theta_{\rm 3dB} =65 ^{\circ}
\end{align}

\begin{figure}[t]
	\includegraphics[scale=0.15]{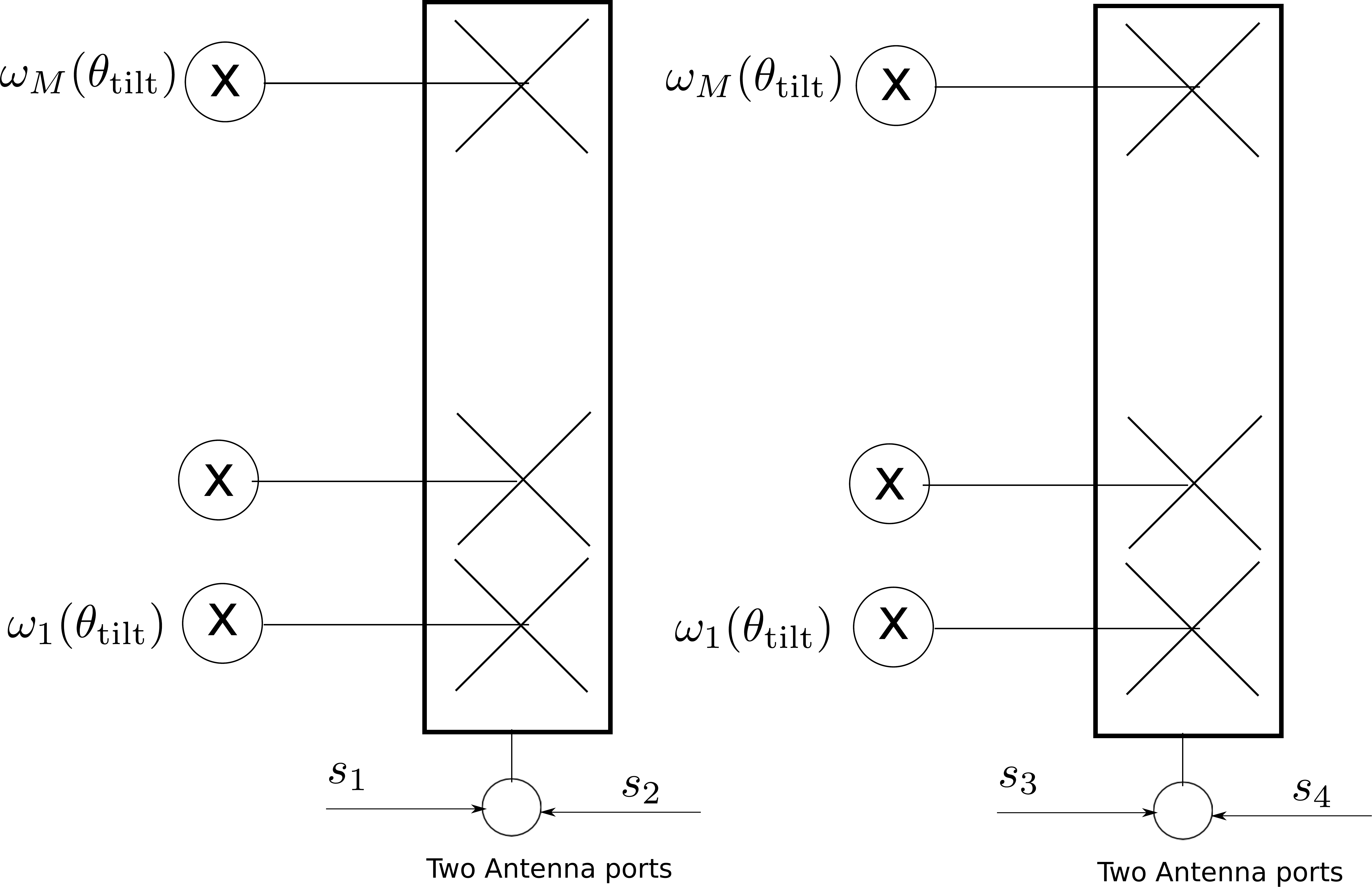}
	\caption{$K=M$}
	\label{fig:KequalM}
\end{figure}

\begin{figure}[t]
\centering{
	\includegraphics[scale=0.15]{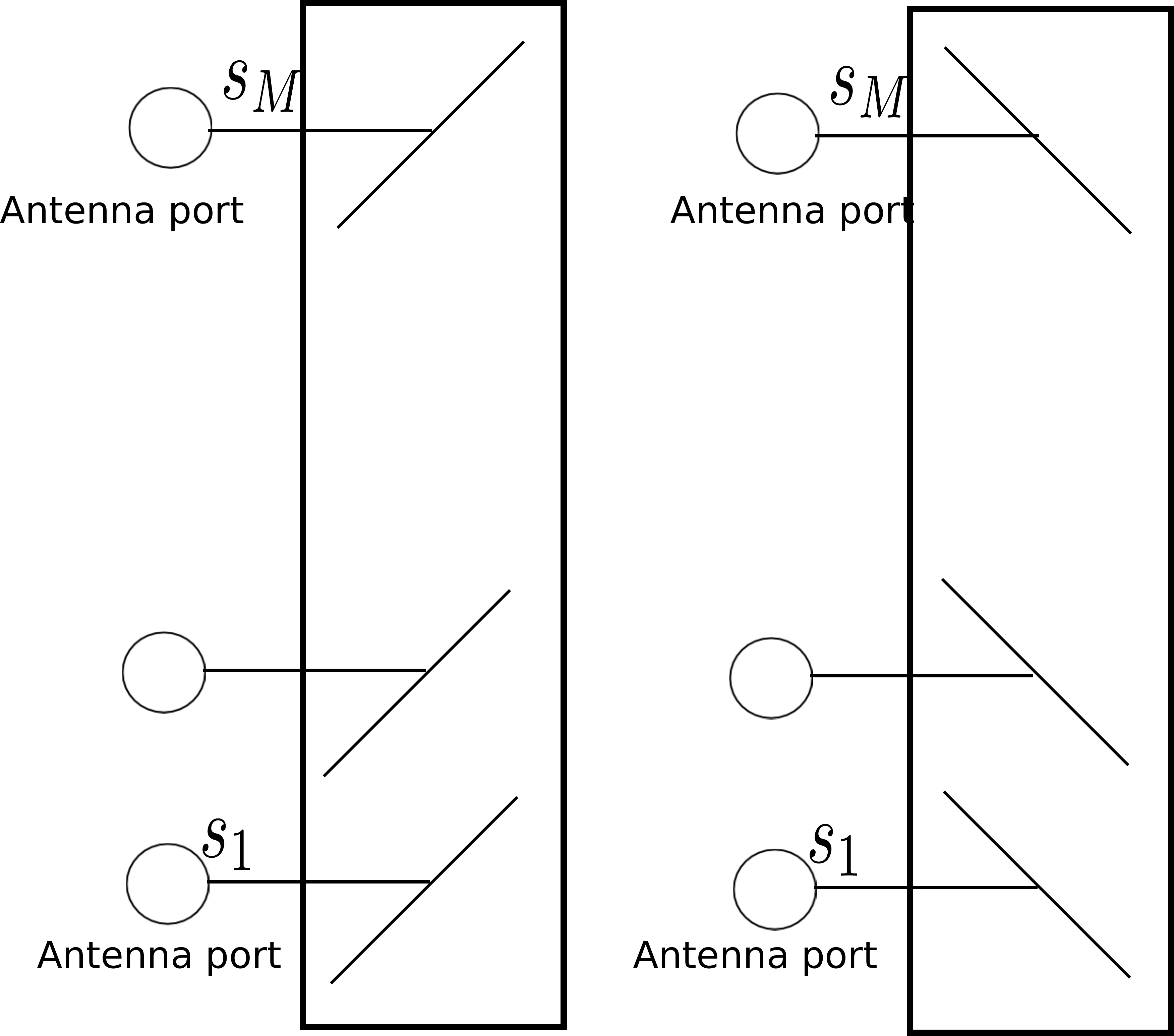}}
	\caption{$K=1$}
	\label{fig:K1}
\end{figure}

Two cases are considered by the 3GPP group TSG-RAN-WG1, either choosing $K=1$ in which case the number of antenna ports per column will be equal to $M$ or setting $K=M$, in which case, each column will correspond to one port with the same polarization (see Fig. \ref{fig:KequalM} and Fig. \ref{fig:K1}). Note that, if cross-polarized elements were used, each column would correspond to two ports, (one port per polarization).
%At the receiver side, each antenna port will appear as a single antenna.
The channel with a given  antenna port is a weighted sum of  channels with  the $K$ antenna elements  inside it. More formally,   the channel
 between the $u$th receiving antenna and the $s$th antenna port corresponding to
 the $n$th path is given by:
% Let $\left[{\bf H}_n\right]_{k,u}$ denote the channel between the $k$-th antenna element and the $u$-th receiving antenna. 

%If we use the expression of the antenna pattern of one element, \eqref{eq:los} will give the channel with transmitting antenna elements. The equivalent channel between the $u$th receiving antenna and the $s$th antenna port corresponding to the $n$th path is given by:
 \begin{equation}
\left[\overline{\bf H}_n\right]_{s,u} = \sum_{k \in \  \textnormal{port} \  s } \omega_k \left[{\bf H}_n\right]_{k,u}
\end{equation}
where the sum above is performed over all antenna elements in port $s$ and ${\bf H}_n$ is computed by using in \eqref{eq:los} the field patterns of one antenna element. This field pattern could be easily retrieved by determining the coordinate transformation between the local  and the global spherical coordinate system, \cite{orange_contrib_pol} \footnote{The local spherical coordinate system is defined by the antenna axis $\vec{\bf e}_{za}$, i.e, $\vec{\bf e}_\phi=\frac{\vec{\bf e}_r\times \vec{\bf e}_{za}}{\|\vec{\bf e}_r\times \vec{\bf e}_{za}\|}$ and $\vec{\bf e}_\theta=\vec{\bf e}_\phi\times \vec{\bf e}_r$, where $\times$ being the cross-product.}.  

The objective of the  channel representation based on the antenna elements is two folds: First it enables to take into account the pattern of  side lobes, an effect which has been discarded by the channel representation in ITU \cite{itu} and 3GPP 36.814 \cite{36814}. And second, it allows more flexibility, the channel being linearly dependent on the weights $\omega_k$. We can also add a third advantage which is the possibility of providing accurate expressions for the vertical and horizontal field patterns.
This has been  the subject of our contribution \cite{orange_contrib_pol} which was prepared for meeting RAN1$\#74$ but  only off-line submitted. In meeting $\#75$, the same polarization model  was proposed by NSN, Nokia\cite{R1-136021}, Qualcomm \cite{R1-135312} and Fraunhofer \cite{R1-135708} and was recently adopted in the last version of the TR36.873 \cite{36873}. %(mets le en reference avaec la date du 2014-02) pour montrer qu'on a mis à jour les informations 
\subsection{System level simulations}
\label{sec:system}
\begin{figure}
	\includegraphics[width=\linewidth]{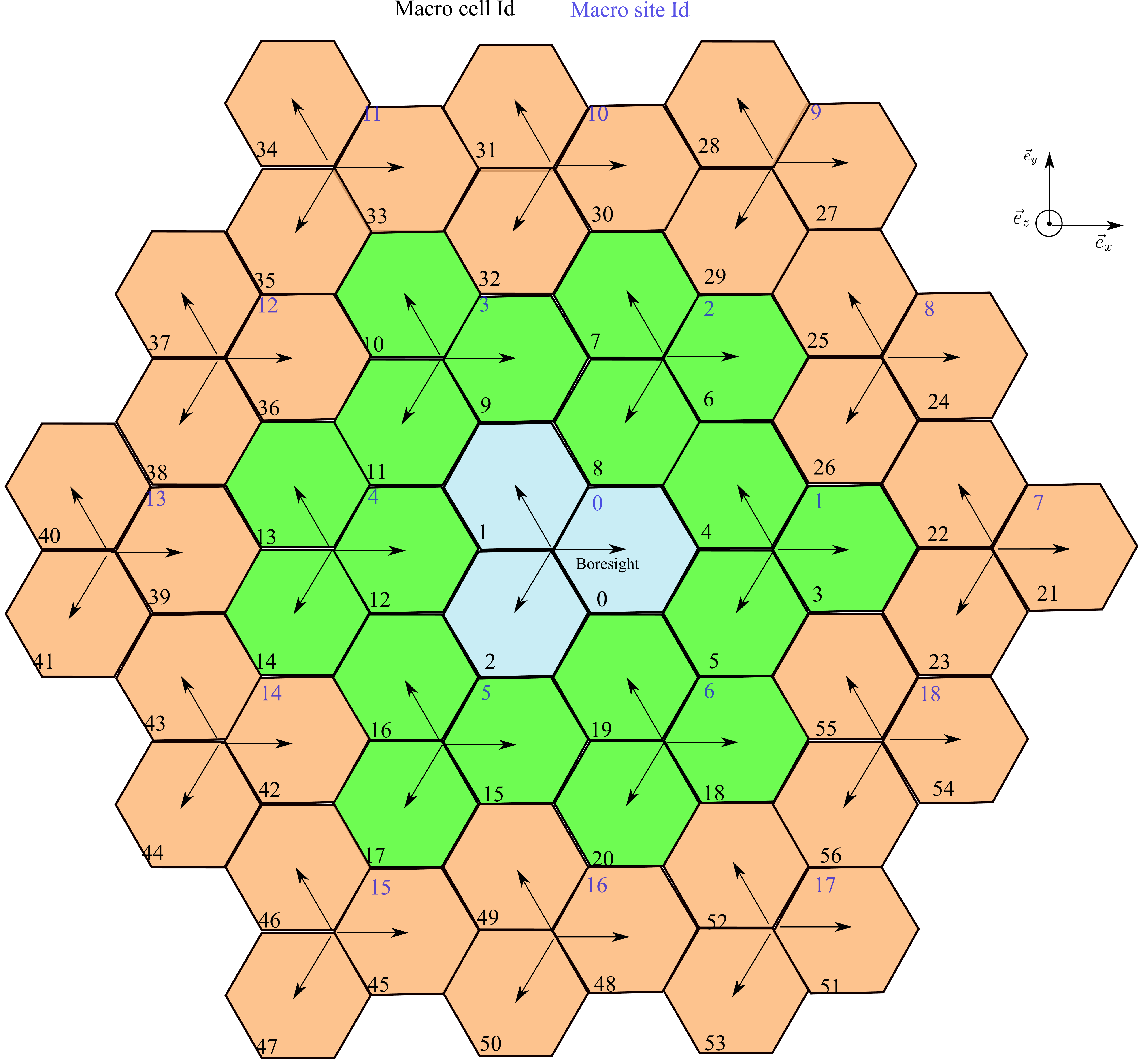}
	\caption{Network Layout}
	\label{fig:network}
\end{figure}

The new channel model that is now being prepared by 3GPP will serve to perform system level simulations.  A network of 19 sites is considered, where each site is composed of  tri-sector base stations (See Fig. \ref{fig:network}). Each base station is equipped with a 2D antenna array structure composed of $M$ rows and $N$ columns.  The TSG-RAN-WG1 is considering the scenario of Urban-Micro (UMi) and Urban-Macro (UMa) as well as an additional scenario describing Urban-Macro environments with high-rise buildings (UMA-H). Measurement campaigns are currently being performed in order to tune the large scale parameters, as well as, the slow fading. An important progress has been made. An agreement has been already reached for the slow fading parameters. Also, the tuning of the large scale parameters is at a final stage. The document  \cite{R1-134980} contains all the decisions  as well as some of the few remaining issues. 

For the reader convenience, we summarize in the next section the new properties which make the difference with previous channel models. 
\subsubsection{Major changes from $2D$ to $3D$}

\begin{enumerate}[a)]
	\item 3D UE dropping: In previous standardized channel models, all the UE are assumed to have a height of $1.5m$. With such UE dropping, the elevation of the LOS direction depends on the distance between the BS and the UE: the closest is the UE to the BS, the largest is the elevation corresponding to the LOS direction. As such, users undergoing similar slow fading channel conditions will show almost the same elevation in the LOS direction,  making it difficult to perform  vertical separation. In light of this consideration, the TSG-RAN WG1 is working  on 3D UE Dropping. With this new dropping, each user is outdoors in only $20\%$ of the cases, in which situation its height is set to $1.5m$. If it is indoors, then it is inside a building of $x$ floors where $x$ is a random variable uniformly chosen from the set $\left\{4,5,6,7,8\right\}$. Finally, the UE floor $n_{fl}$ is uniform-randomly selected from the set $\left\{1,\cdots,x\right\} $ while the UE height is given by $h_{UE}=3(n_{fl}-1)+1.5$. 
% we correspond randomly  height ranging from $1.5m$ to $22.5m$, If it is in outdoors, the user height is set to $1.5m$ similarly to the actual existing standards. If it is an indoor user, 
\item Outdoor to Indoor users for UMa and UMi: In the ITU channel, the outdoor-to indoor connection is a propagation mode specific to the Urban-Microcell scenario which was proposed to support UEs inside buildings.  The 3GPP new channel model will extend this propagation mode to the UMa since this scenario will consider also indoor users.   
\item High rise scenario: High rise scenario is an optional scenario proposed by CMCC \cite{cmccR1-133528} in meeting RAN1$\#74$ to describe the channel of UEs in very high floors (above $20$). To cover these users, a possible way is to deploy indoor distributed antenna systems, an alternative which has been already deployed in high office buildings. Unfortunately, this solution cannot be always applied for high residence buildings. In this case, the use of uptilt beams can be proposed.  
\item New expression for the path-loss: 
 It is well known that, in addition to the frequency and the distance with the UE,  the path loss depends also on the heights of transmitting and receiving antennas \cite{ikegami}. A correction factor should be thus introduced in order to account for these dependencies. Based on measurements, the TSG-RAN-WG1 modifies the pathloss expressions of the ITU channel model  in order to account for the impact of the UE height in NLOS conditions. Morevoer, a 3D distance instead of 2D distance is now used in the path-loss expression.  
\item Distance dependent large scale parameters: Very recently, based on measurements, NSN, Nokia proposes to model the elevation spreads as  log-normal random variables whose parameters depend on the 2D distance for  UMa scenarios and on both the  2D distance and the mobile height for UMi scenarios, \cite{R1-135948}. This model has just been approved in meeting RAN1$\#75$.
%\item Polarization: As we mentioned earlier, the 3GPP 36.814 used an improper polarization model. We have proposed in a contribution that we prepared for meeting 74 and that was unfortunately not officially submitted an alternative polarization model which holds as long as the impact of the reflecting plane is negligible.
\item UE Attachment Policy: In system level simulators, the determination of the serving cell for each UE is based on  a metric  known as RSRP (Reference Signal Received Power). Currently, the computation of the RSRP is based solely on the slow fading parameters (Pathloss + shadow fading). More details about the computation of the RSRP in this case is given in section 2.\ref{sec:phase1}) dealing with phase 1 calibration. For phase 2 and phase 3 calibrations, it has just been agreed in meeting  RAN1$\#75$ to account also for  fast fading \cite{R1-136000}. This method should be more accurate, since unlike the azimuth angles, the elevation angles are modeled as random variables whose mean is shifted by a constant angle from the LOS direction. 
	%We were the first to note that and to propose a well-founded polarization model, a work which was prepared for the meeting 74 and that was unfortunately not officially submitted \cite{orange_contrib_pol}. Two meetings after, four companies which are NSN, Nokia, Qualcomm and Fraunhofer IIS  have proposed the same proper modeling. At the time of the writing of this paper, the proposed polarization model is still under discussion.
\end{enumerate}

\subsubsection{Calibration process}
%Channel models that are described in  standards belong to the class of geometry-based stochastic channel models in that they rely on the concept of multi-path clusters where each cluster is a group of unresolvable multipath components sharing similar parameters. Two main approaches have been proposed in standards to parametrize the clusters: a system level approach used by the 3GPP Spatial Channel Model (SCM) \cite{3gpp} and the WINNER II/WINNER + models \cite{winner}
% There are two different for characterizing these clusters: 
Companies and organizations involved in channel standardization conduct measurements campaigns in order to tune the channel system parameters. Once an agreement is reached, companies implement the channel in their own simulation tools, which are not publicly available. Although being based on the same channel model, companies come in general to different system performances. The reason can be attributed to either a mismatch on scheduling algorithms which are not standardized, or to some errors in the code, a problem which is likely to often occur given the high complexity of the simulation procedure. To exclude the second reason, channel calibration is often performed. For calibrations that do not consider a transmission scheme, most of the companies  have to get similar results. The TSG-RAN WG1 considers three calibration phases. We present hereafter these calibrations as well as the obtained so far results. 
%At the time of writing this paper, calibrations are still being conducted by the TSG-RAN WG1. 
%In this section, we present the different calibrations considered in the  TSG-RAN WG1. 
\begin{enumerate}[a)]
\item{Phase 1 Calibration:}
\label{sec:phase1}
Phase 1 calibration relies only on the slow fading parameters. The objective of this phase is to compute the empirical cumulative distribution function (CDF) of two metrics referred to as coupling gain and geometry factor. The coupling gain measures the difference between the received signal and the transmitted signal along the line-of sight direction.  On the other hand, the geometry factor measures the signal to interference ratio with respect to the serving cell. At the time of the writing of the paper, TSG-RAN WG1 has just finished the calibration phase 1. We will present in this section our results which comply with those obtained so far by the 3GPP TSG-RAN-WG1.  But, before that,   we detail hereafter for the reader convenience the procedure that has been followed to achieve the phase 1 calibration. 

After performing UE dropping, this phase consists in computing for each UE  $m$ and site $k$ the following quantities:
\begin{itemize}
	\item  The LOS direction characterized by a spatial angle $\Omega_{k,m}=(\phi_{LOS,k,m},\theta_{LOS,k,m})$ and $\Psi_k=\left(\varphi_{LOS,k,m},\vartheta_{LOS,k,m}\right)$ with respect to the $k$-th cell,
	\item For each cell $i$ in the site ($i=1,2,3$), compute the antenna gain at the transmitter and the receiver $G_T(\phi_{LOS,k,m,i},\theta_{LOS,k,m,i})$ and $G_R(\varphi_{LOS,k,m,i},\vartheta_{LOS,k,m,i})$ in ${\rm dB}$,
	\item Compute the pathLoss $PL_{k,m}$ and shadow-fading $\sigma_{{SF}_{k,m}}$ between the UE and cell $k$,
	\item Define for the user, the $RSRP_{k,m,i}$ (Reference Signal Received Power)
		\begin{align}
			RSRP_{k,m,i}&= P_{TX} + G_T(\phi_{LOS,k,m,i},\theta_{LOS,k,m,i}) \nonumber \\
								&+ G_R(\phi_{LOS,k,m,i},\theta_{LOS ,k,m,i})\nonumber\\
					   &- PL_{k,m}({\rm dB}) -\sigma_{{SF},k,m}({\rm dB})
	\end{align}
	where $P_{TX}$ is the transmitted power in ${\rm dB}$. 
	\item Determine the index $\overline{k}_{m}$ and $\overline{i}_{m}$ of the best serving cell which maximizes the RSRP.  
	\item The coupling gain for the UE $m$ is defined as:
		\begin{equation}
		CL_m= RSRP_{\overline{k}_{m},\overline{i}_{m}} -P_{TX}
	\end{equation}
	\item The geometry factor for the UE $m$ is given by:
		\begin{equation}
		GF= RSRP_{\overline{k}_{m},\overline{i}_{m},m}-\sum_{(k,i)\neq (\overline{k}_{m},\overline{i}_m)}RSRP_{k,i,m}
	\end{equation}
\end{itemize}
We present in Fig. \ref{fig:dv5} and Fig. \ref{fig:dv8}, the obtained results of the geometry factor for the UMa scenario when the vertical spacing is equal to $0.5\lambda$ and $0.8\lambda$. 
These results comply with those obtained by majority of the companies participating in the calibration process. Note that for $d_v=0.5 \lambda$ the  downtilt of  $12$ degree achieves the best performance. This is to be compared to the case where $d_v=0.8\lambda$  for which the best dowtilt value is given by $9$ degree. In this case, the reduction of inter-cell interference for high downtilt angles ($12$ degree) cannot  compensate the degradation in  coupling gain which is amplified by the fact that the beam is narrower. 
\begin{figure}[t]
\begin{minipage}[c]{0.5\textwidth}
\includegraphics[scale=0.4]{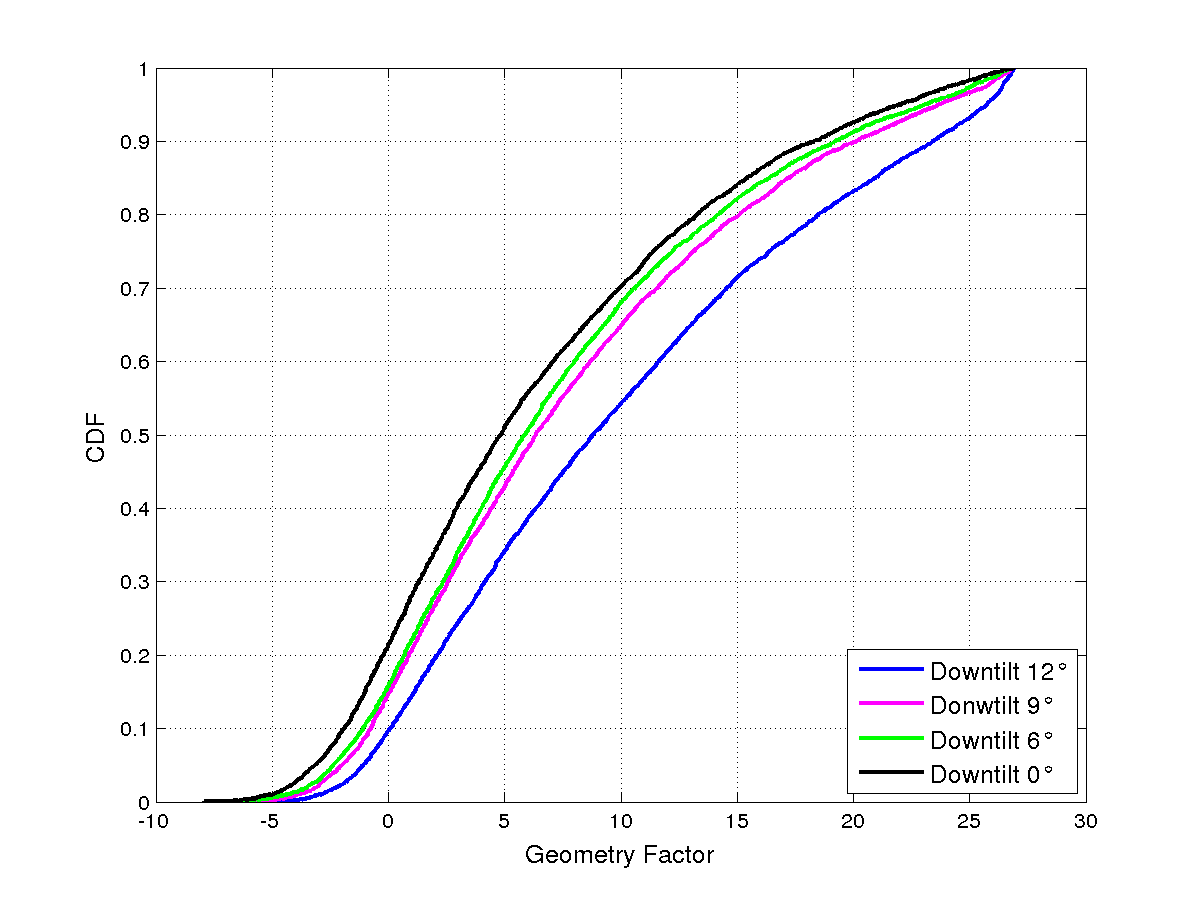}
\caption{Geometry factor for $d_v=0.5\lambda$}
\label{fig:dv5}
\end{minipage}
\begin{minipage}[c]{0.5\textwidth}
\includegraphics[scale=0.4]{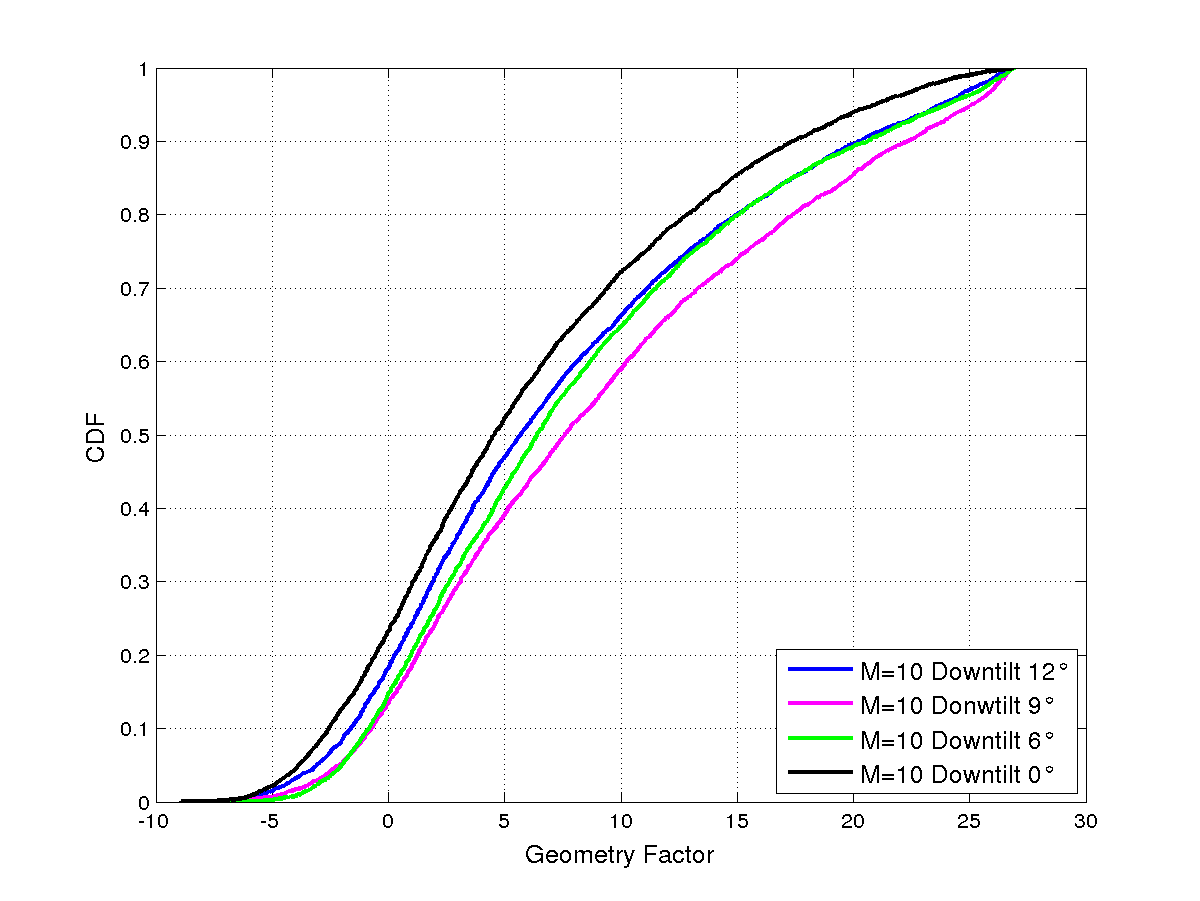}
\caption{Geometry factor for $d_v=0.8\lambda$}
 \label{fig:dv8}
\end{minipage}
\end{figure}
Finally, for sake of comparison with respect to the previous ITU 2D channel model, we superpose in fig.\ref{fig:2dvs3d} the geometry factor obtained for the 3D channel when $d_v=0.5\lambda$ with that of the ITU-2D channel model \cite{itu}. Note that the 3D-3GPP channel model achieve better performance in terms of geometry factor. This result is not surprising since the 3D channel model employ a 3D UE dropping and as such UEs in higher floors undergo lower interference from other cells. The implementation of future 3D and FD MIMO techniques over 3D channels should allow to evaluate in a more accurate way their real performances.
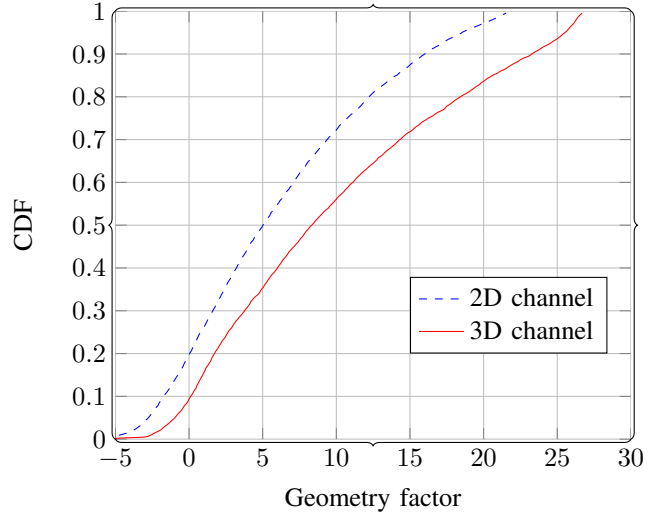
\begin{figure}[h]
	\begin{tikzpicture}[scale=1]
		%\path[red] (axis cs: 1,0) -- (axis cs: 1,0.6);
	%\path[red] (axis cs: 2,0) -- (axis cs: 2,0.6);
	%\path[red] (axis cs: 7,0) -- (axis cs: 7,0.6);
		
	\begin{axis}[xmajorgrids,ymajorgrids, xlabel=Geometry factor, ylabel=CDF,xmin=-5,ymin=0,
xmax=30,
ymax=1, enlarge x limits=false,
ytick={0,0.1,0.2,0.3,0.4,0.5,0.6,0.7,0.8,0.9,1},
legend style={
	{nodes=right},
	anchor=south west,
	at={(axis cs:15,0.2)}
}
]
%\addplot[red,very thick]{};
\addplot[smooth, dashed,blue,no markers] table[each nth point=50, x=g1, y=g2, col sep=space]{file_2d.dat};
\addplot[smooth, red,no markers] table[each nth point=50, x=g1, y=g2, col sep=space]{file_3d.dat};
%\addplot[smooth, red, very thick] gnuplot [raw gnuplot] {plot "file_2d.dat" using ($0*1000):1 every 1000};
%\addplot[smooth, blue, very thick] file {file_3d.dat};
%\addplot[red,very thick] plot coordinates{(0.010000,0.000268)(0.020000,0.000277)(0.030000,0.000286)(0.040000,0.000295)(0.050000,0.000305)(0.060000,0.000315)(0.070000,0.000326)(0.080000,0.000338)(0.090000,0.000350)(0.100000,0.000364)(0.110000,0.000378)(0.120000,0.000393)(0.130000,0.000409)(0.140000,0.000427)(0.150000,0.000446)(0.160000,0.000466)(0.170000,0.000488)(0.180000,0.000512)(0.190000,0.000538)(0.200000,0.000566)(0.210000,0.000597)(0.220000,0.000632)(0.230000,0.000670)(0.240000,0.000713)(0.250000,0.000760)};
\addlegendentry{2D channel }
\addlegendentry{3D channel}
%\addlegendentry{Mesure th\'eorique}
\end{axis}
\end{tikzpicture}
\caption{Comparison with the performance of the ITU channel model\cite{itu}}
\label{fig:2dvs3d}
\end{figure}
\item{Phase 2 and Phase 3 Calibrations:}
Phase 2 and Phase 3 calibrations are now being conducted by the TSG-RAN-WG1.  
As we mentioned above, for phase 2 and 3 calibrations, the computation of the RSRP takes into account the fast fading. For each $m$ UE and site $k$, large scale parameters describing the distribution of the azimuth and elevation angles are generated. With these large scale parameters on hand, small scale parameters including azimuth and elevation angles, powers and delays are generated for each UE and cells. Note that the links between a UE and cells in the same site share the same large scale parameters but different small scale parameters. Unlike phase 1 calibration, the RSRP for all the link is computed based on the fast fading in addition to the slow fading. The UE is then associated to the base station which delivers the highest RSRP. 

For phase 2 calibration, companies should give CDF curves for the azimuth and elevation angular spreads as well as for the delay spread. To calibrate the channel generation procedure,  CDF of the first and second largest eigenvalues of the channel should also be presented \cite{R1-136000}. 

Finally for phase 3 calibration, a transmission scheme will be considered, and the system level performances will be expressed in terms cell coverage, cell edge throughput and spectrum efficiency, \cite{R1-134069}. 
\end{enumerate}
\section{Conclusion}

 Channel modeling is sparking increasing interest from both academia and industry. A number of different approaches in standards and theory have arisen. Nevertheless, few works have considered to establish a common thread between them. This has been in particular the objective of our work. After a brief overview of the theoretical basics of channel modeling, we have presented some of the most used standardized channel models, with a special focus on their different and common properties. A close inspection of these models reveals that they  have evolved towards a higher complexity as they should satisfy more stringent requirements.
 With the emergence of a growing interest for elevation beamforming, a large effort is now being devoted to produce accurate 3D channel models. This will enable in future a fair evaluation of 3D beamforming techniques. The 3GPP TSG-RAN-WG1 group is now working on producing  3D channel models that will be be of fundamental practical use. Several new strategies have been agreed so far within the 3GPP. These aspects being unknown, we presented in this paper a brief overview of the main features of the proposed channel and conducted some simulations in order to compare its performance with that of the 2D standard channel model.
 %To pave the way towards 3D beamforming techniques, these models should account for the channel elevation component. 
 % At the moment, a large effort is being made within the 3GPP TSG-RAN-WG1 group to produce accurate 3D channel models, thereby paving the way towards 3D beamforming techniques. 
 %In this line, we have provided in this paper a brief overview on the 3D channel modeling that is being carried out  by the 3GPP, as well as some initial results concerning phase 1 calibration.

 \section{Acknowledgements}
This research is supported by the french p\^ole de comp\'etitivit\'e SYSTEM@TIC within the project 4G in Vitro.
%In this paper, we have provided a survey on 3D channel modeling in recent standards. A special focus has been made on system approach based standards 
%This paper provides a survey on 3D channel modeling in recent standards. After a brief review of the theoretical basics of channel modeling, we have presented some of the most used standardized channel models. 
\bibliographystyle{IEEEbib}
\bibliography{IEEEabrv,IEEEconf,tutorial_RMT}

\begin{thebibliography}{10}

\bibitem{hata}
M.~Hata,
\newblock ``{Empirical Formulas for Propagation Loss in Land Mobile Radio
  Service},''
\newblock {\em {IEEE} Trans. Veh. Technol.}, vol. 29, no. 3, pp. 317--325, Aug.
  1980.

\bibitem{lee}
W.~C.~Y. Lee,
\newblock {\em {Mobile Communications Engineering}},
\newblock McGrawHill, New York, 1982.

\bibitem{jakes-71}
W.~C. Jakes,
\newblock ``{A Comparison of Specific Space Diversity Techniques for Reduction
  of Fast Fading in UHF Mobile Radio Systems},''
\newblock {\em {IEEE} Trans. Veh. Technol.}, vol. 20, no. 4, pp. 81--91, Nov.
  1971.

\bibitem{Failli}
M.~Failli,
\newblock ``{Digital Land Mobile Communications COST 207 Final Report},''
\newblock {\em {Office for Official Publications of the European Communities,
  Final report}}, 1989.

\bibitem{Damosso}
E.~Damosso and L.~M. Correia,
\newblock ``{Digital Mobile Radio Towards Future Generation Systems},''
\newblock {\em Final Report of COST Action 231, European Union}, 1999.

\bibitem{Correia}
L.~M. Correia,
\newblock {\em {Wireless Flexible Personalized Communications COST 259:European
  Co-operation in Mobile Radio Research }},
\newblock John Wiley \& Sons, New York, 2001.

\bibitem{3gpp}
``{Spatial Channel Model for Multiple Input Multiple Output (MIMO)
  Simulations},'' \url{http://www.3gpp.org/ftp/Specs/html-info/25996.htm},
  Sept. 2003.

\bibitem{extended-3gpp}
D.~S. Baum, J.~Hansen, and J.~Salo,
\newblock ``{An Interim Channel Model for Beyond 3G Systems : Extending the
  3GPP Spatial Channel Model (SCM)},''
\newblock in {\em Proc. {IEEE} Vehicular Technology Conference}, May 2005,
  vol.~5, pp. 3132--3136.

\bibitem{winner}
IST-4-027756~WINNER II,
\newblock ``{D1.1.2, WINNER II Channel Models},''
  \url{https://www.ist-winner.org/WINNER2-Deliverables/D1.1.2v1.1.pdf}, Sept.
  2007.

\bibitem{itu}
Report ITU-R M.2135,
\newblock ``{Guidelines for evaluation of radio interface technologies for
  IMT-advanced},'' \url{http://www.itu.int/pub/R-REP-M.2135-2008/en}, 2008.

\bibitem{koppenborg}
J.~Koppenborg, H.~Halbauer, S.~Saur, and C.~Hoek,
\newblock ``{3D Beamforming Trials with an Active Antenna Array},''
\newblock in {\em ITG Workshop on Smart Antennas}, 2012.

\bibitem{3gpp-study}
R1-122034,
\newblock ``{Study on 3D channel Model for elevation Beamforming and FD-MIMO
  studies for LTE},''
\newblock {\em 3GPP TSG RAN Plenary \# 58}, Dec. 2012.

\bibitem{rappaport}
T.~Rappaport,
\newblock {\em {Wireless Communications, Principles and Practice}},
\newblock Prentice-Hall, Englewood Cliffs, NJ, USA, 1996.

\bibitem{molish-book}
A.~F. Molisch,
\newblock {\em {Wireless Communications}},
\newblock Wiley IEEE Press, New York,NY, USA, 2005.

\bibitem{correai}
L.~Correia,
\newblock {\em {Mobile Broadband Multimedia Networks}},
\newblock John Wiley \& Sons, New York, NY, USA, 2006.

\bibitem{bruno}
B.~Clerckx and C.~Oestges,
\newblock {\em {MIMO Wireless Networks : Channels, Techniques and Standards for
  Multi-antenna, Multi-User and Multi-Cell Systems}},
\newblock Academic Press (Elsevier), Oxford, UK, Jan. 2013.

\bibitem{R1-133719}
R1-133719,
\newblock ``{Initial Results for 3D channel Model UMa Calibration Case 1,
  Orange},''
\newblock {\em 3GPP TSG RAN1 Meeting \#74}, Aug. 2013.

\bibitem{R1-133720}
R1-133720,
\newblock ``{Initial Results for 3D channel Model UMa Calibration Case 3,
  Orange},''
\newblock {\em 3GPP TSG RAN1 Meeting \#74}, Aug. 2013.

\bibitem{pahlavan}
K.~Pahlavan and A.~H. Levesque,
\newblock {\em {Wireless Information Networks}},
\newblock John Wiley \& Sons, New York, 1995.

\bibitem{petrus02}
P.~Petrus, J.~H. Reed, and T.~S. Rappaport,
\newblock ``{Geometrical-Based Statistical Macro cell Channel Model for Mobile
  environments},''
\newblock {\em {IEEE} Trans. Commun.}, vol. 50, no. 3, March 2002.

\bibitem{chen03}
Y.~Chen and V.~K. Dubey,
\newblock ``{A Generic Channel Model in Multi-cluster environments},''
\newblock in {\em VTC}, Apr. 2003, vol.~1, pp. 217--221.

\bibitem{cardoso-1}
F.~D. Cardoso and L.~M. Correia,
\newblock ``{A Time-Domain Based Approach for Short-term Fading Depth
  Evaluation in Wideband Mobile Communication Systems},''
\newblock {\em Wireless Personal Communications}, vol. 35, no. 4, pp. 365--381,
  Dec. 2005.

\bibitem{cardoso-2}
F.~D. Cardoso and L.~M. Correia,
\newblock ``{Fading Depth Dependence on System BandWidth in Mobile
  Communications: An Analytical Approximation},''
\newblock {\em {IEEE} Trans. Veh. Technol.}, vol. 52, no. 3, pp. 587--594, May
  2003.

\bibitem{steinbauer}
M.~Steinbauer, A.~F. Molisch, and E.~Bonek,
\newblock ``{The Double-Directional Radio Channel},''
\newblock {\em {IEEE} Antennas Propag. Mag.}, vol. 43, pp. 51--63, Aug. 2001.

\bibitem{molisch-04}
A.~F. Molisch,
\newblock ``{A Generic Model For MIMO Wireless Propagation Channels in Macro-
  and Microcells},''
\newblock {\em {IEEE} Trans. Signal Process.}, vol. 52, no. 1, Jan. 2004.

\bibitem{molish}
A.~F. Molisch, H.~Asplund, R.~Heddergott, M.~Steinbauer, and T.~Zwick,
\newblock ``{The COST 259 Directional Channel Model-Part I: Overview and
  Methodology},''
\newblock {\em {IEEE} Trans. Wireless Commun.}, vol. 5, no. 12, pp. 3421--3433,
  2006.

\bibitem{D5.3}
J.~Meinil\"a, P.~Ky\"osti, L.~Hentil\"a, T.~J\"ams"a, E.~Suikkanen, E.~Kunnari,
  and M.~Narand\v{z}i\'c,
\newblock ``{D5.3: WINNER+ Final Channel Models},'' June 2010.

\bibitem{burbank}
J.~L. Burbank, J.~Andrusenko, J.~S. Everett, and W.~T.~M. Kasch,
\newblock {\em Wireless Networking Understanding Internetworking Challenges},
\newblock IEEE Press, Piscataway, New Jersey, 2013.

\bibitem{36814}
3GPP TR 36.814~V9.00 3rd,
\newblock {\em Further advancements for E-UTRA physical layer aspects}, Mar.
  2010.

\bibitem{37840}
3GPP TR~37.840 V12.00,
\newblock {\em Study of Radio Frequency (RF) and Electromagnetic Compatibility
  (EMC) requirements for Active Antenna Array System (AAS) base station}, Mar.
  2013.

\bibitem{orange_contrib_pol}
R1-xxxxxx,
\newblock ``{Polarization Model for 3D Channel, Orange},''
\newblock {\em Submitted offline,
  \url{http://www.flexible-radio.com/sites/default/files/attachments-186/Polarization%20Model%20For%203D%20Channel.pdf}},
  Aug. 2013.

\bibitem{R1-136021}
R1-136021,
\newblock ``{TP for Mechanical Tilt},''
\newblock {\em NSN, Nokia, 3GPP RAN WG1 Meeting \#75}, Nov. 2011.

\bibitem{R1-135312}
R1-135312,
\newblock ``{Field pattern model of polarized antennas},''
\newblock {\em Qualcomm, 3GPP RAN WG1 Meeting \#75}, Nov. 2011.

\bibitem{R1-135708}
R1-135708,
\newblock ``{Text Proposal on Mechanical and Electrical Antenna Tilting},''
\newblock {\em Fraunhofer IIS, 3GPP RAN WG1 Meeting \#75}, Nov. 2011.

\bibitem{36873}
3GPP TR~36.873 R1-141062,
\newblock {\em Study on 3D channel model for LTE}, Feb. 2014.

\bibitem{R1-134980}
R1-134980,
\newblock {\em {Study on 3D channel model for LTE (Release 12)}}, Sept. 2013.

\bibitem{cmccR1-133528}
R1-133528,
\newblock ``{On Optional High Rise Scenario},''
\newblock {\em 3GPP TSG-RAN WG1 Meeting \#74}, Aug. 2013.

\bibitem{ikegami}
F.~Ikegami, S.~Yoshida, T.~Takeuchi, and M.~Umehira,
\newblock ``{Propagation Factors Controlling Mean Field Strengthb on Urban
  Streets},''
\newblock {\em {IEEE} Trans. Antennas Propag.}, vol. 32, pp. 822--829, 1984.

\bibitem{R1-135948}
R1-135948,
\newblock ``{Remaining details of fast-fading for 3D-UMa and 3D-UMa},'' Nov.
  2013.

\bibitem{R1-136000}
R1-136000,
\newblock ``{WF on RSRP calculation formula},''
\newblock {\em 3GPP TSG RAN WG1 Meeting \#75}, Nov. 2013.

\bibitem{R1-134069}
R1-134069,
\newblock ``{Simulation assumptions for calibration and baseline performance
  evaluation},''
\newblock {\em Huawei, HiSilicon, 3GPP TSG RAN WG1 Meeting \#75}, Nov. 2013.

\end{thebibliography}
\begin{IEEEbiographynophoto}
{Abla Kammoun} was born in Sfax, Tunisia. She received the engineering degree in signal and systems from the Tunisia Polytechnic School, La Marsa, and the Master's degree and the Ph.D. degree in digital communications from Telecom Paris Tech [then Ecole Nationale Supérieure des Télécommunications (ENST)]. From June 2010 to April 2012, she has been a Postdoctoral Researcher in the TSI Department, Telecom Paris Tech. Then she has been at Supélec at the Alcatel-Lucent Chair on Flexible Radio until December 2013. Currently, she is a Postodoctoral fellow at KAUST university. Her research interests include performance analysis, random matrix theory, and semi-blind channel estimation.
\end{IEEEbiographynophoto}
\begin{IEEEbiographynophoto}
{Dr. Hajer Khanfir} received the B.S. degree in Communication Technologies from SUP'COM in 1996; M.S. from the Ecole Nationale d'Ingénieurs de Tunis (ENIT) in 2002 and PhD degree from Conservatoire National des Arts et Métiers of Paris in 2009. She has worked for 7 years in Tunisie Telecom as wireless network engineer. Since 2009 she has joined Orange Labs France where she is working on various PHY and PHY-MAC in 4G mobile radio techniques and is also a 3GPP standardization delegate.
\end{IEEEbiographynophoto}
\begin{IEEEbiographynophoto}
{Dr. Zwi Altman} received the B.Sc. and M.Sc. in the Technion-Israel Institute of Technology, in 1986 and 1989 respectively, and the Ph.D. From the INPT France in 1994. From 1994 to 1996 he was a Post-Doctoral Research Fellow in the University of Illinois at Urbana Champaign. He joined Orange Labs in 1996 where he has been involved in various projects on network planning and optimization, automatic cell planning, self-organizing networks, autonomics and performance evaluation. Dr. Altman is a senior research expert in Orange.
\end{IEEEbiographynophoto}
\begin{IEEEbiographynophoto}
{M\'erouane Debbah} entered the Ecole Normale Supérieure de Cachan (France) in 1996 where he received his M.Sc and Ph.D. degrees respectively. He worked for Motorola Labs (Saclay, France) from 1999-2002 and the Vienna Research Center for Telecommunications (Vienna, Austria) until 2003. He then joined the Mobile Communications department of the Institut Eurecom (Sophia Antipolis, France) as an Assistant Professor until 2007. He is now a Full Professor at Supelec (Gif-sur-Yvette, France), holder of the Alcatel-Lucent Chair on Flexible Radio and a recipient of the ERC grant MORE (Advanced Mathematical Tools for Complex Network Engineering). His research interests are in information theory, signal processing and wireless communications. He is a senior area editor for IEEE Transactions on Signal Processing and an Associate Editor in Chief of the journal Random Matrix: Theory and Applications. Mérouane Debbah is the recipient of the "Mario Boella" award in 2005, the 2007 General Symposium IEEE GLOBECOM best paper award, the Wi-Opt 2009 best paper award, the 2010 Newcom++ best paper award, the WUN CogCom Best Paper 2012 and 2013 Award, the 2014 WCNC best paper award as well as the Valuetools 2007, Valuetools 2008, CrownCom2009 and Valuetools 2012 best student paper awards. In 2011, he received the IEEE Glavieux Prize Award and in 2012, the Qualcomm Innovation Prize Award. He is a WWRF fellow and a member of the academic senate of Paris-Saclay.
\end{IEEEbiographynophoto}
\begin{IEEEbiographynophoto}{Mohamed Kamoun}  received the engineering degree from
Ecole National Supérieure de Techniques Avancées, Paris, France in 2001, a master degree in semi-conductor physics from the Université Paris Sud, Orsay, France, in 2001, a master degree in digital communication from Ecole Nationale Supérieure des Télécommunications in 2002, and the Ph.D. degree from the Université Paris sud, Orsay, France in 2006. From 2002 to 2008 he has been working for Motorola Laboratories, Paris. Since 2009 he works in the LSC laboratory of CEA-List as researcher. His research interest include cooperative communications, network coding, and next generation cellular networks.
\end{IEEEbiographynophoto}

\end{document}